\documentclass{sig-alternate-2013}
\usepackage{graphicx}
\usepackage{mathtools,breqn}
\usepackage{tabularx,booktabs}
\usepackage{array}
\usepackage{dcolumn}
\usepackage{tabu}
\usepackage{algorithm}
\usepackage[noend]{algpseudocode}
\usepackage{amsfonts}
\usepackage{subfigure}
\usepackage{epsfig}
\newcolumntype{F}{>{\bfseries}c}

\def\vio{\mathcal{V}}
\def\nonvio{\mathcal{S}}
\def\gang{\textit{gang}}
\def\arr{\textit{arr}}
\def\T{\textsf{true}}
\def\F{\textsf{false}}
\def\C{\textsf{C}}
\def\P{\textsf{P}}
\def\distr{\textit{distr}}
\def\beat{\textit{beat}}
\def\tm{t_{\max}}
\def\st{\textit{ s.t. }}
\def\naturals{\mathbb{N}}

\makeatletter
\def\@copyrightspace{\relax}
\makeatother

\begin{document}

\title{Early Identification of Violent Criminal Gang Members}
\numberofauthors{2} 
\author{
\alignauthor
Elham Shaabani\thanks{\small{These authors contributed equally to this work.}} , Ashkan Aleali$^*$, and Paulo Shakarian\thanks{\small{U.S. Provisional Patent 62/191,086. Contact shak@asu.edu for licensing information.}}\\\smallskip
       \affaddr{Arizona State University}\\
       \affaddr{Tempe, AZ}\\
       \email{\{shaabani, aleali, shak\}@asu.edu} 
\alignauthor John Bertetto\\\smallskip
       \affaddr{Chicago Police Department}\\
       \affaddr{Chicago, IL}\\
       \email{john.bertetto@chicagopolice.org}
}
\maketitle

\begin{abstract}
Gang violence is a major problem in the United States accounting for a large fraction of homicides and other violent crime.  In this paper, we study the problem of early identification of violent gang members.  Our approach relies on modified centrality measures that take into account additional data of the individuals in the social network of co-arrestees which together with other arrest metadata provide a rich set of features for a classification algorithm.  We show our approach obtains high precision and recall (0.89 and 0.78 respectively) in the case where the entire network is known and out-performs current approaches used by law-enforcement to the problem in the case where the network is discovered overtime by virtue of new arrests - mimicking real-world law-enforcement operations.  Operational issues are also discussed as we are preparing to leverage this method in an operational environment.
\end{abstract}

\vspace{1mm}
\noindent
{\bf Categories and Subject Descriptors:} J.4 {[Computer Applications]}: {Sociology}

\vspace{1mm}
\noindent
{\bf General Terms:} Security; Experimentation

\vspace{1mm}

\section{Introduction}

Gang violence is a major problem in the United States~\cite{bertetto12,braga13} - accounting for 20 to 50 percent of homicides in many major cities~\cite{howell}.  Yet, law enforcement actually has existing data on many of these groups.  For example the underlying social network structure is often recorded by law-enforcement and has previously been shown useful in enabling ``smart policing'' tactics~\cite{pap12} and improving law\\-enforcement's understanding of a gang's organizational structure~\cite{damon13}.  In this paper we look to leverage this gang social network information to create features that allows us to classify individuals as potentially violent.  While the results of such a classifier are insufficient to lead to arrests, it is able to provide the police leads to individuals who are likely to be involved in violence, allowing for a more focused policing with respect to patrols and intelligence gathering.  \textit{Our key aim is to significantly reduce the population of potential violent gang members which will lead to more efficient policing.}

In this paper, we introduce our method for identifying potentially violent gang members that leverages features derived from the co-arestee social network of criminal gangs in a classifier to identify potentially violent individuals.  We note that this classification problem is particularly difficult due to not only data imbalances, but also due to the fact that many violent crimes are conducted due to heightened emotions - and hence difficult to identify.  Though we augment our network-based features with some additional meta-data from the arrest records, our approach does \textbf{not} leverage features concerning the race, ethnicity, or gender of individuals in the social network.  We evaluate our method using real-world offender data from the Chicago Police Department.  This paper makes the following contributions:

\begin{itemize}

	\item{We discuss how centrality measurements such as degree, closeness, and betweenness when modified to account for metadata about past offenses such as the type of offense and whether the offense was classified as ``violent'' can serve as robust features for identifying violent offenders.}
	
	\item{We show how the network features, combined with other feature categories provide surprisingly robust performance when the entire offender is known in terms of both precision (0.89) and recall (0.78) using cross-fold validation.}
	
	\item{We then test our methods in the case where the network is exposed over time (by virtue of new arrests) which mimics an operational situation.  Though precision and recall are reduced in this case, we show that our method significantly outperforms the baseline approach currently in use by law-enforcement - on average increasing precision and recall by more than two and three times respectively.}

\end{itemize}

In addition to these main results, we also present some side results on the structure and nature of the police dataset we examine.  The paper is organized as follows.  In Section~\ref{relvSec} we motivate this difficult problem within the law-enforcement community.  This is followed by a description of our dataset along with technical notation in Section~\ref{sec:prelim}.  There, we also describe some interesting aspects of the gang arrest dataset and our co-arrestee network.  In Section~\ref{vpredSec} we formally define our problem, describe existing approaches, and then describe the features we use in our approach.  Then we present our results in Section~\ref{sec:exp_res} for both cases where we assume the underlying network is known and when we discover the network over time (mimicking an operational scenario).  Finally, related work is discussed in Section~\ref{rwSec}.

\section{Background}
\label{relvSec}
A recent study shows that the network for gunshot victimization is denser than previously believed~\cite{pap15}. According to the authors, within the city of Chicago over 70\% of all gunshot victims are contained within only 6\% of the total population These findings validate what has been considered common knowledge among police for decades: who you hang out with matters, and if you hang out with those who engage in or are victims of violence you are more likely to become an offender or victim yourself.\smallskip

Identifying potential offenders of gun violence has also been a staple practice for most law enforcement agencies as an attempt to curtail future victimization. When gang conflicts get ``hot,'' it's common for law enforcement agents to put together a list of known ``shooters'': those known gang members with an existing criminal history for gun violence and a predilection for engaging in such illegal activity. Law enforcement agents then attempt to make contact with these individuals with the expectation that such direct contact might prevent violence. For most law enforcement agencies, however, this practice is performed in a very ad-hoc manner. Identifying these individuals for intervention has relied primarily on the ability of law enforcement agents to remember and identify at-risk individuals. While feasible for small or discreet networks, the ability to recall multiple individuals in large networks that cross large geographic regions and interact with multiple networks becomes increasing difficult. This difficulty increases significantly as relationships between networks change, known individuals leave the network, and new individuals enter it. In particular, the practice is less than idea because it requires officers to attempt to recall criminal history and network association data that varies between network members. For example, a subject who has been arrested on multiple occasions for carrying a gun or has been arrested for shooting another individual is easy to recall, but recalling and quantifying the risk for a subject with multiple arrests for non-gun violence and a direct association with several offenders and victims of gun violence can be much more difficult. In short, identifying a known ``shooter'' is relatively straightforward: they are known. The approach in this paper synthesizes network connectivity other attributes of the subject to identify those individuals at risk that law enforcement might not yet know.\smallskip

Using this information, law enforcement agents may not only more reliably and consistently identify those individuals most likely to engage in acts of violence or become victims of violence due to their personal associations with it, but also to more effectively manage agency resources. Intervention strategies may include service providers outside law enforcement, such as family members, social service providers, current or former educators, and clergy. This diversity in approach not only delivers a more powerful ``stop the violence'' message but provides a kind of force multiplier for law enforcement, increasing the number of persons involved in the effort to prevent violence. Identifying specific individuals for intervention also allows for a more targeted effort by law enforcement in terms of personnel and geographic areas needing coverage. Blanketing violence reduction strategies that saturate geographic areas with law enforcement agents and rely on direct contact with large numbers of criminal network members are inefficient and resource consuming. Focusing efforts on those individuals most likely to engage in violence allows law enforcement to focus on smaller groups of people and smaller geographic areas (those areas within which those individuals identified are known to frequent).  Therefore, our approach can significantly improve such efforts to identify violent individuals.  In this paper, we see how our method not only out-performs the current social network heuristic used by police, but also that it provides a much smaller and more precise list of potentially violent offenders than simply listing those with a violent criminal record.

\section{Gang Co-Offender Network}
\label{sec:prelim}
In this section, we introduce the necessary basic notation to describe our co-offender network and then provide details of our real-world criminal dataset and study some of its properties.

\subsection{Technical Preliminaries}
Throughout this paper we shall represent an offender network as an undirected graph $G=(V, E)$ where the nodes correspond with previous offenders and an undirected edge exists between offenders if they were arrested together.  We will use $\tau$ to denote the set of timepoints (dates).  We also have three sets of labels for the nodes: $\vio,\ \nonvio,\ \gang$ which are the sets of violent crimes, non violent crimes, and gangs.  For each time point $t$ and each node $v$, the binary variable $\arr_v^t \in \{\T,\F\}$ denotes if $v$ was arrested at time $t$ and $\distr_v^t,\beat_v^t,\gang_v^t$ to denote the district, beat, and gang affiliation of $v$ at time $t$ (we will assume that time is fine-grain enough to ensure that at each time unit an individual is arrested no more than once).  If we drop the $t$ superscript for these three symbols, it will denote the most recent district, beat, and gang associated with $v$ in the knowledgebase.  We shall use the sets $\vio_v^t$ and $\nonvio_v^t$ to denote the set of violent and non violent offenses committed by $v$ at time $t$ respectively.  Note if $\arr_v^t=\F$ then $\vio_v^t=\emptyset$.  We will drop the superscript $t$ for this symbol to denote the union of labels at any time $t$ in the historical knowledgebase.  We also note that the edges in the graph also depend on time, but for sake of readability, we shall state with words the duration of time considered for the edges.\smallskip

For a given violent crime $c\in \vio \cup \nonvio$, we will use the notation $V_{c}^t=\{v \in V \textit{ s.t. } c \in \vio_v^t\}$ (intuitively, the subset of the population who have committed crime $c$ at time $t$).  Again, we will drop the superscript $t$ if $v$ could have committed crime $c$ at any time in the historical knowledgebase.  For a set of labels $C \subseteq \vio \cup \nonvio$, we will extend this notation: $V_C^t=\{v \in V \textit{ s.t. } C \cap \vio_v^t \neq \emptyset\}$.  We will slightly abuse notation here: $V_{\emptyset}^t = V$.  We will use similar notation for denoting a subset of the population that are members of a certain gang.  For instance, $V_{\gang_v}$ refers to the set of nodes who are in the same gang as node $v$.  Likewise, we shall use the same notation for subgraphs: $G_C^t$ is the subgraph of $G$ containing only nodes in $V_C^t$ and their adjacent edges.  We will use the function $d: V \times V \rightarrow \naturals$ to denote the distance between two nodes - which for this paper will be the number of links in the shortest path.  For a given node $v$, the set $N_v^i=\{v' \in V \textit{ s.t. }d(v,v')=i\}$ -- the set of nodes that are whose shortest path is exactly $i$ hops from $v$.  For two nodes $v,v'$, we will use the notation $\sigma(v,v')$ to be the number of shortest paths between $v$ and $v'$.  For nodes $u,v,v'$, $\sigma_u(v,v')$ will be the number of shortest paths between $v$ and $v'$ that pass through $u$.\smallskip

For a given subgraph $G'$ of $G$, we shall use $\C(G')$ to denote the largest connected component of $G'$ and for node $v \in G'$, we will use the notation $\C_v(G')$ to denote the connected component of $G'$ to which $v$ belongs.  If we apply a community finding algorithm to subgraph $G'$, we will use the notation $\P_v(G')$ to denote the partition of $G'$ to which $v$ belongs.  We will use the notation $|\cdot|$ to denote the size of a set or the number of nodes in a subgraph.

\subsection{Overview of Network Data}
In this section we describe our police dataset and the associated co-offender network as well as some interesting characteristics that we have noticed.\smallskip\\

\noindent{\textbf{Police Dataset.}}  Our dataset consists of gang-related arrest incidents gathered from August 2011 - August 2014 in Chicago as well as their immediate associates. This data set includes locations, dates, the links between the joint arrests, and the gang affiliation of the offenders.  In Table~\ref{arSumTab}, we summarize some of the important characteristics of the dataset.\smallskip\\

\begin{table}
\centering
\caption{\textmd{Summary of arrest data.}}
		\begin{tabular}{| l | c|}
			\hline
			\textbf{Name} & \textbf{Value}\\\hline \hline
			Number of records & 64466\\ \hline
			Violent offense & 4450\\
			\ \ \ Homicide & 312\\
			\ \ \ Criminal sexual assault & 153\\
			\ \ \ Robbery & 1959\\
			\ \ \ Aggravated assault & 1441\\
			\ \ \ Aggravated battery & 896\\ \hline
			Non violent offense & 60016\\ \hline
		\end{tabular}
	\label{arSumTab}
\end{table}

\begin{table}
	\centering
	\caption{\textmd{Network properties.}}
	\begin{tabular}{| p{5cm}| c|}
		\hline
		\textbf{Name} & \textbf{Values} \\ \hline \hline
		Vertices & 9373 \\ \hline
		Edges & 17197 \\ \hline
		Average degree & 3.66 \\ \hline
		Average clustering & 0.5 \\ \hline
		Transitivity & 0.62 \\ \hline
		Connected components & 1843 \\ \hline
		Largest connected component diameter & 36 \\ \hline
		Largest connected component average path length & 12.22 \\ \hline
		Largest connected component average clustering & 0.63 \\ \hline
	\end{tabular}
	\label{tab:net_prop}
\end{table}

\begin{figure}[h!]
	\centering
	\includegraphics[width=7cm]{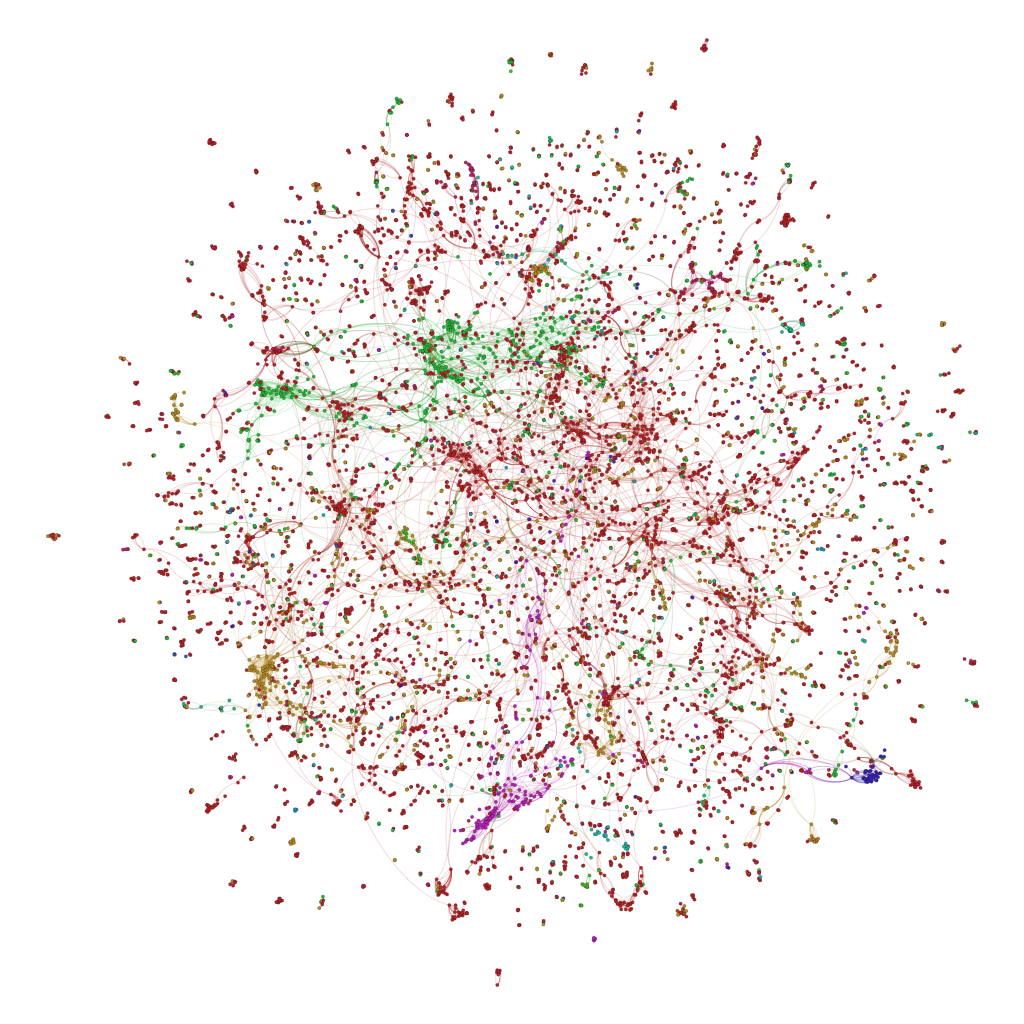}
	\caption{\textmd{The gang co-offender network. Each color corresponds with a different gang.}}
	\label{fig:graph_example}
\end{figure}

\noindent{\textbf{Violent Crimes.}}  In our dataset, the set $\vio$ consists of the following crimes have been identified by the Chicago Police as violent crimes: homicide (first or second degree murder),  criminal sexual assault, robbery, aggravated assault, and aggravated battery.  All aforementioned offenses are also FBI ``index'' crimes as well.  A key aspect about the violent crimes is that the dataset is highly imbalanced with much more arrests for non violent crimes vs. arrests for violent crimes (60016 vs. 4450).\smallskip\\

\begin{figure}[h!]
	\centering
	\epsfig{figure=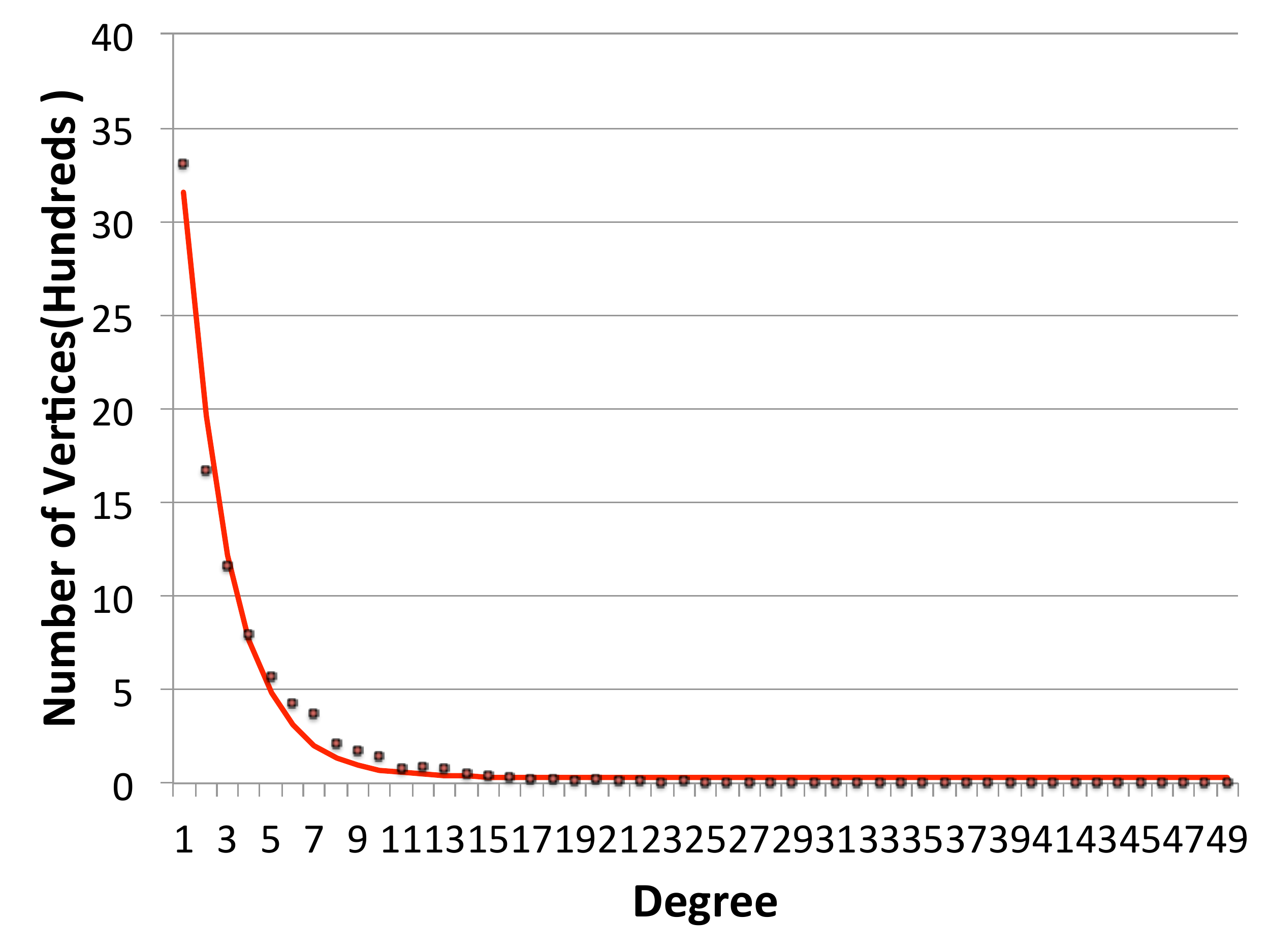,width=7cm}
	\caption{\textmd{Network degree distribution. The exponential function fits to the distribution ($R^2=0.77$).}}
	\label{fig:degree_dist}
\end{figure}

\noindent{\textbf{Network Properties.}}  From the arrest data, we were able to construct the \emph{co-offender network}.  In this network, the isolated vertices are eliminated  due to the lack of structural information.  A visualization of the network is depicted in Figure~\ref{fig:graph_example} and we have included summary statistics in Table~\ref{tab:net_prop}.  In studying this network, we studied its degree distribution (Figure~\ref{fig:degree_dist}). Unlike the degree distribution for other scale free social networks, the degree distribution for the offender network is \emph{exponential} rather than \emph{power law}. However, despite the degree distribution being similar to that of a random (E-R) or small world network topology~\cite{watts1998collective}, we noticed other characteristics that indicate differently.  The co-offender network has a much higher average clustering coefficient than in a random network and does not follow the properties of the small world topology due to the relative high diameter and average shortest path (computed for the largest connected component.)\smallskip\\

\begin{figure}[h!]
	\centering
	\epsfig{figure=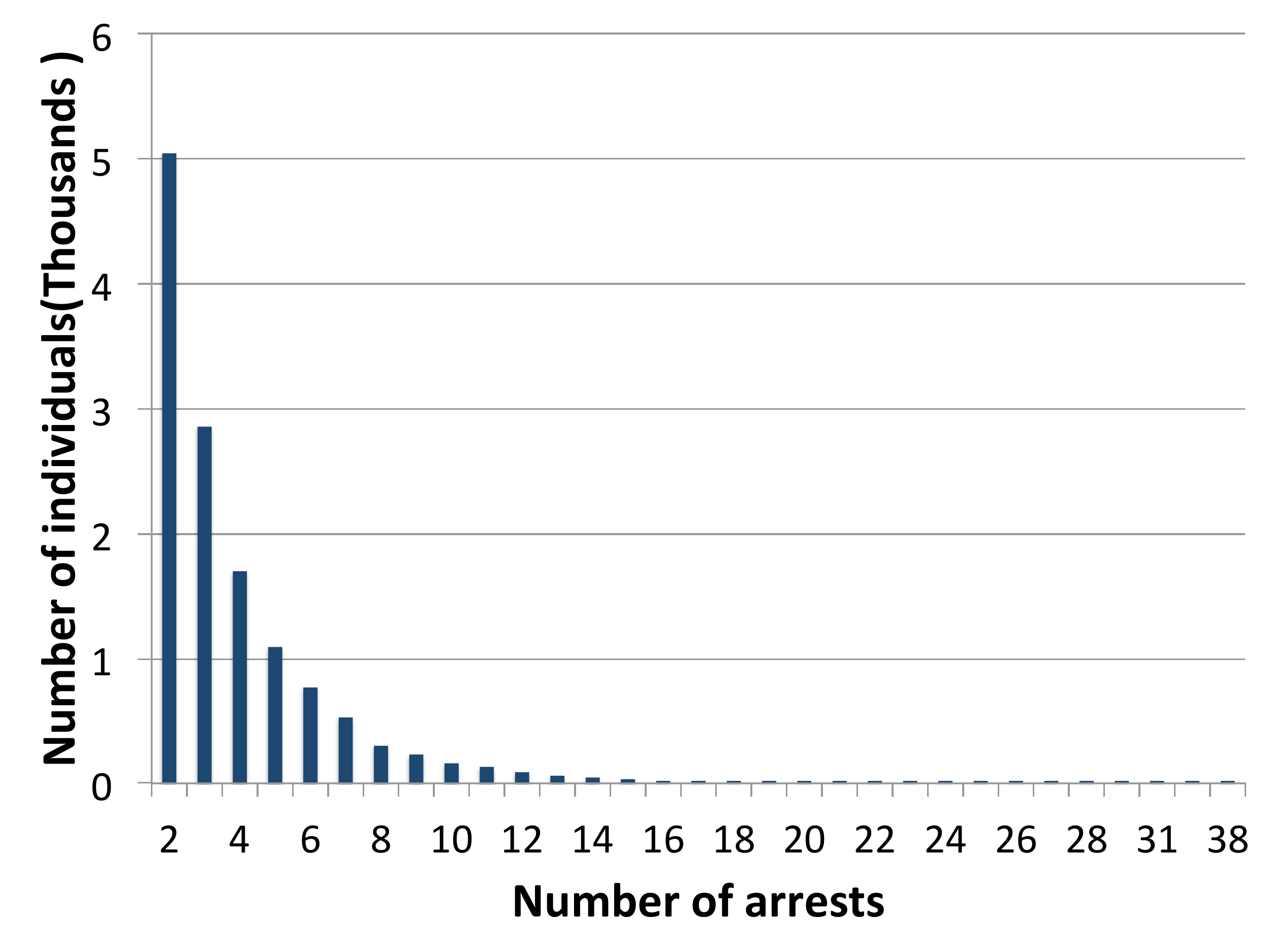,width=7cm}
	\caption{\textmd{Repeated arrests. 12866 instances of one-time arrests have been removed.}}
	\label{fig:rpt} 
\end{figure}

\noindent{\textbf{Repeat Offenders.}} There are many instances of repeated offenses from the same offender. Figure~\ref{fig:rpt} shows the distribution of the repeated arrests for each individual in the dataset.  This indicates that arrest records have utility in identifying future offenders.\smallskip\\

\begin{figure}[h!]
	\centering
		\subfigure {
			\epsfig{figure=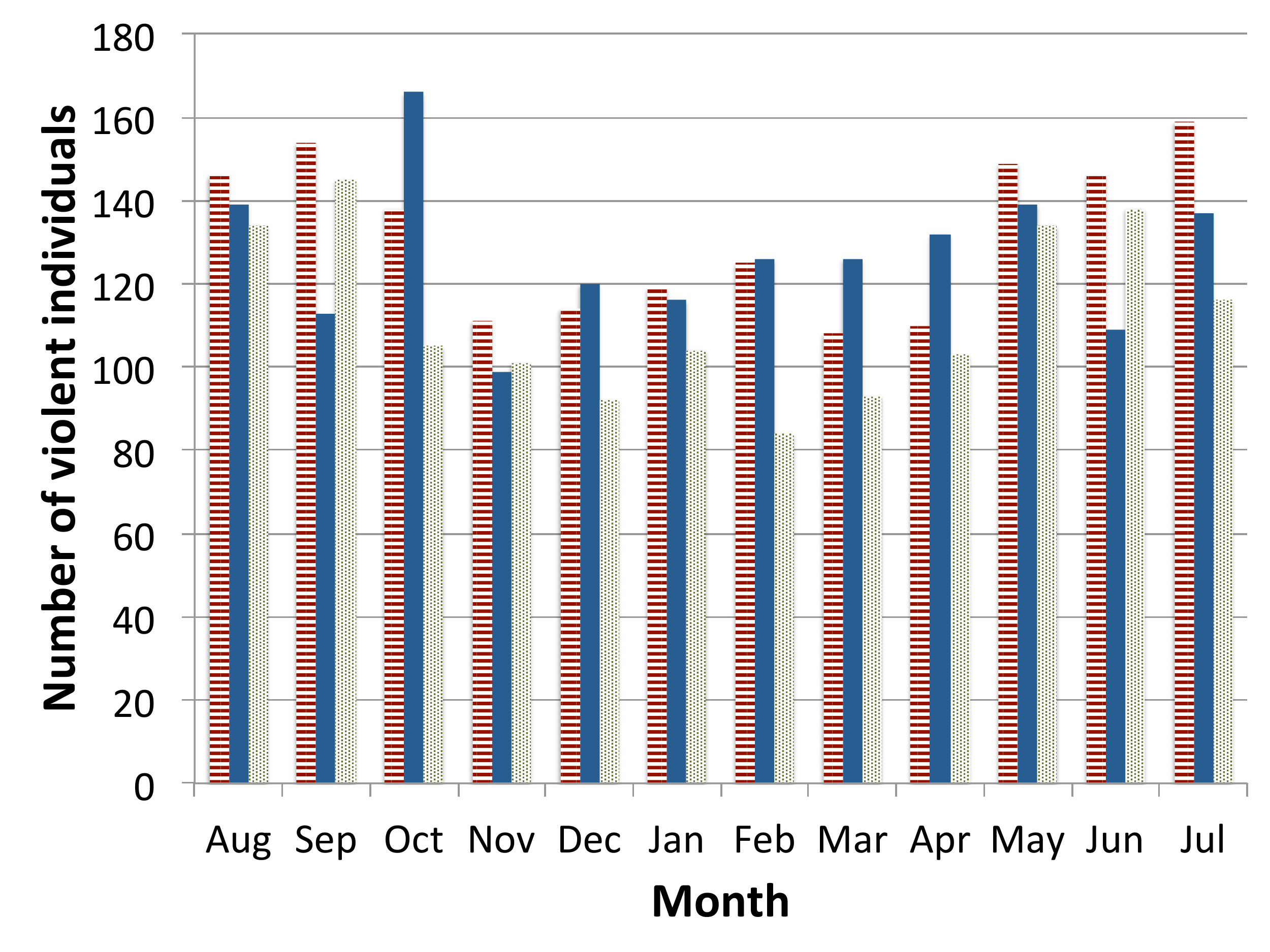,width=7cm}
		}
		\quad
		\subfigure {
			\epsfig{figure=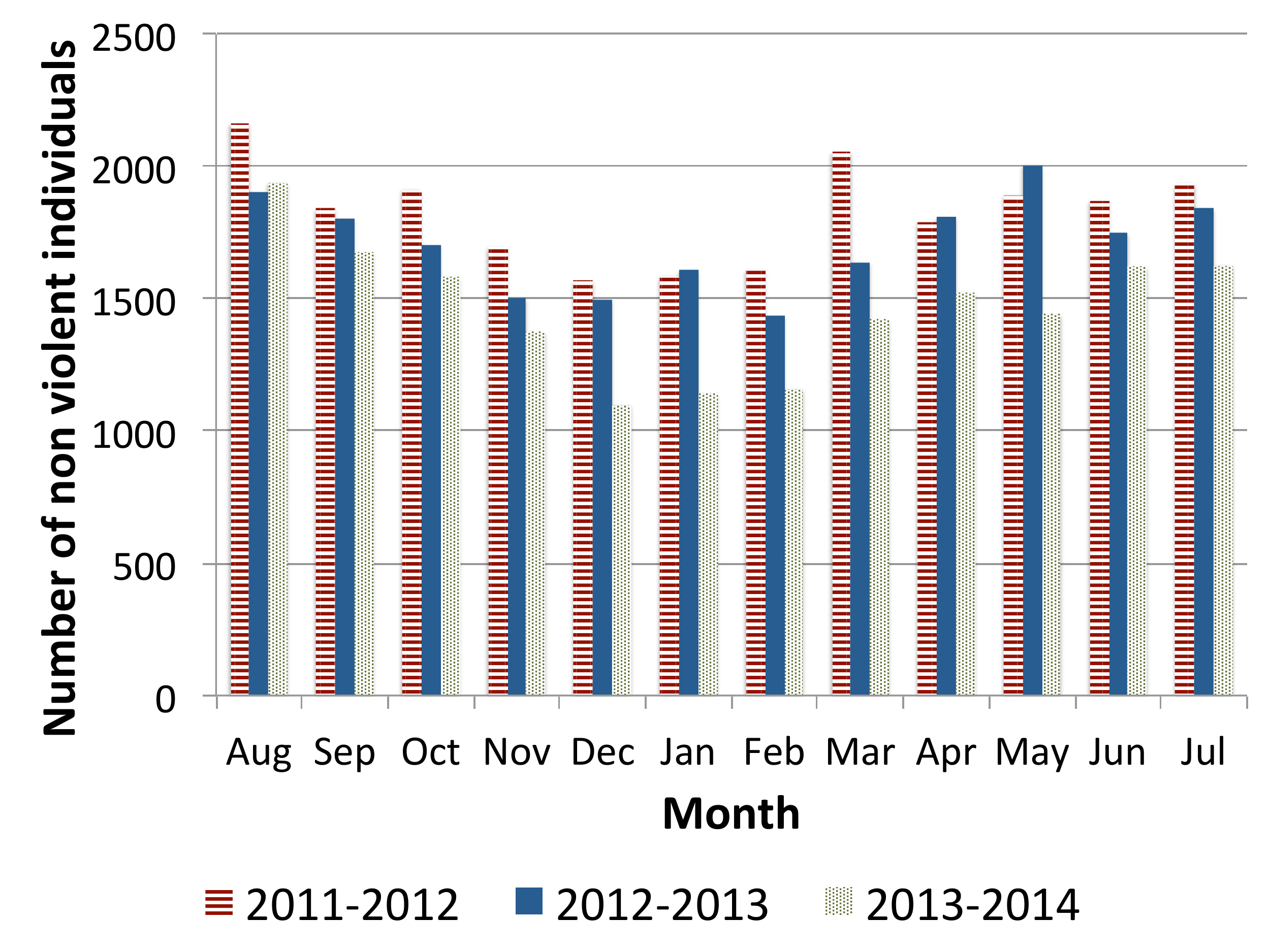,width=7cm}
		}
		\quad
		\caption{\textmd{Seasonality of crime.}}
		\label{fig:seasonality}
\end{figure}

\noindent{\textbf{Seasonality of Crime.}} There is also a higher chance of criminal activities in different months of the year. Figure~\ref{fig:seasonality} demonstrates some of these variations.  As per police observations, both violent and non-violent crime incidents are lower in the winter months (Dec.-Feb.).\smallskip

\section{Identifying Violent Offenders}
\label{vpredSec}
In this section, we describe our problem, some of the existing practical approaches used by law-enforcement, and our approach based on supervised learning with features primarily generated by the network topology.

\subsection{Problem Statement}
Given a co-offender network, $G=(V,E)$ and for each historical timepoint $t \in \tau=\{1,\ldots,\tm\}$ and $v \in V$, we have the values of $\arr_v^t, \distr_v^t,\beat_v^t$ and elements of the sets $\vio_v^t,\gang_v^t$, we wish to identify set $\{ v\in V \textit{ s.t. } \exists t >\tm \textit{ where }|\vio_v^t| >0 \}$.  In other words, we wish to find a set of offenders in our current co-offender network that commit a violent crime in the future.

\subsection{Existing Methods}
Here we describe two common techniques often used by law-enforcement to predict violent offenders.  The first is a simple heuristic based on violent activities in the past.  The second is a heuristic that was based on the findings of \cite{pap12} which was designed to locate future \textit{victims} of violent crime.  Both of these approaches are ad-hoc practical approaches that have become ``best practices'' for predicting violent offenders.  However, we are not aware of any data-driven, formal evaluation of these methods in the literature.\smallskip\\

\noindent\textbf{Past Violent Activities (PVA).}  The first ad-hoc approach is quite simple: if an offender has committed a violent crime in the past, we claim that he will commit a violent crime in the future.  An obvious variant of this approach is to return the set of violent offenders from the last $\Delta t$ days.  We note in practice, if police also have records of those who are incarcerated, and such individuals would be removed from the list (due to the different jurisdictions of police and corrections in the Chicago area, we did not have access to incarceration data - however discussed re-arrests observed in the data in the previous section).\smallskip\\

\noindent\textbf{Two-Hop Heuristic (THH).}  The two-hop heuristic is based on the result of \cite{pap12} which investigated a social network of gunshot \textit{victims} in Boston and found an inverse relationship between the probability of being a gunshot victim and the shortest path distance on the network to the nearest previous gunshot victim.  Hence, THH returns all neighbors one and two hops away from previous violent criminals (see Algorithm~\ref{alg:two_hops} for details on the version we used in our experiments - which was the best-performing variant for our data).  The Chicago Police have adopted a variant of this method to identify potential gang victims using a combination of arrest and victim data - the co-arrestee network of criminal gang members includes many individuals who are also victims of violent crime (this is a direct result of gang conflict).  We note that victim information did not offer a significant improvement to our approach, except the trivial case that a homicide victim cannot commit any crime in the future.\smallskip

\begin{algorithm}
	\caption{Two-Hop Heuristic}\label{alg:two_hops}
	\begin{algorithmic}[1]
		\Procedure{TwoHop}{$G$}\Comment{Offender network $G$.}
		\State $R \gets \{\}$\Comment{Identified violent offenders.}
		\State $VICTIMS \gets \{ u \in G | is\_homicide\_victim(u) \}$
		\For{$v \in VICTIMS$}
		\State $N \gets N_v^1 \cup N_v^2$\Comment{Immediate neighbors}
		\State $R \gets R \cup \{u \in N\ s.t.\ \vio_u = \emptyset\}$
		\EndFor
		\State \textbf{return} $R$
		\EndProcedure
	\end{algorithmic}
\end{algorithm}

\subsection{Supervised Learning Approach}
\label{sec:approach}
We evaluated many different supervised learning approaches including Naive Bayes (NB), Linear Regression (LR), Decision Tree (DT), Random Forest (RF), Neural Network (NN), and support vector machines (SVM) on the same set of features for the nodes in the network that we shall describe in this section.  We also explored combining these approaches with techniques for imbalanced data such as SMOTE~\cite{chawla2002smote} and Borderline SMOTE~\cite{han2005borderline}, however we do not report the results of Borderline SMOTE as it provided no significant difference from SMOTE.  We group our features into four categories: (1.) neighborhood-based (having to do with the immediate neighbors of a given node), (2.) network-based (features that require the consideration of more than a nodes immediate and nearby neighbors), (3.) temporal characteristics, and (4.) geographic characteristics.

\subsubsection{Neighborhood-Based Features}

Neighborhood-based features are the features computed using each node and its first and/or second level neighbors in $G$ -- often with respect to some $C \subseteq \vio$.  The simplest such measure is the degree of vertex $v$ -- corresponding to the number of offenders arrested with $v$.  We can easily extend this for some set of crimes of interest ($C$) where we look at all the neighbors of $v$ who have committed a crime in $C$.  This generalizes degree (as that is the case where $C=\emptyset$).  In our experiments, we found the most useful neighborhood features to be in the case where $C = \vio$ though standard degree ($C = \emptyset$) was also used.  We also found that using combinations of the following booleans based on the below definition also proved to be useful: 

\begin{eqnarray*}
maj_v(C,i) &=& |\{ u | u \in (\cup_i N_v^i) \cap V_C\}| \geq 0.5 \times |(\cup_i N_v^i)|
\end{eqnarray*}

Intuitively, $maj_v(C,i)$ is $\T$ if at least half of the nodes within a network distance of $i$ from node $v$ have committed a crime in $C$ and $\F$ otherwise.  Using these intuitions, we explored the space of variants of these neighborhood-based features and list those we found to be best-performing in Table~\ref{neighFeat}.

\begin{table}[h!]
	\caption{\textmd{Neighborhood-Based Features}}
	\label{neighFeat}
	\centering
	\renewcommand{\arraystretch}{1.5}
	
	\begin{tabular}{|p{2.5cm}|p{4.5cm}|} 
		\hline
		{\bf Description} &  {\bf Definition} \\ \hline \hline
		Degree (w.r.t. $C$) & $|\{ u | u \in N_v^1 \cap V_C \}|$ \\  \hline
			
		Fraction of 1-hop neighbors committing a crime in $C$ & $ |\{u|u \in N_v^1 \cap V_C\}|/|N_v^1| $ \\\hline
			
		Fraction of 2-hop neighbors committing a crime in $C$ &$|\{u| u \in N_v^2\cap V_C\}|/|N_v^2| $ \\ 	\hline			
				
		Majority of 1-hop and 2-hop neighbors committing a crime in $C$ & 	$maj_v(C,1)\wedge maj_v(C,2)$\\\hline
				
		Minority of 1-hop and majority of 2-hop neighbors comitting a crime in $C$ & $\neg maj_v(C,1)\wedge maj_v(C,2)$\\\hline
	\end{tabular}
	
\end{table}

\begin{table}[h!]
	\caption{\textmd{Network-Based Features (Community)}}
	\label{commFeat}
	\centering
	\renewcommand{\arraystretch}{1.5}
	
	\begin{tabular}{|p{2.5cm}|p{4.5cm}|} 
		\hline
		{\bf Description}  & {\bf Definition} \\ \hline \hline
		Component size when $v$ is removed & $|\C(\C_v(G)\setminus\{v\})|$\\  \hline
							
		Largest component size with a violent node after $v$ is removed & $\max_{v' \in \C(\C_v(G)\{v\} \cap V_{\vio}}|X_{v'}|$ where $X_{v'}=\C_{v'}(\C_v(G)\{v\})$\\\hline
			
		Group size & $| \P_v(G_{\gang_v}) |$ \\\hline 

		Relationships within the group & $|\{(u,v)\in E \st u,v \in \P_v(G_{\gang_v}) \}|$ \\\hline

		Number of violent members in the group & $| \{v' \in \P_v(G_{\gang_v})\st \vio_v \neq \emptyset \}|$ \\\hline

		Triangles in group & No. of triangles within subgraph $\P_v(G_{\gang_v})$ \\ \hline
		
		Transitivity of group & $\frac{\textit{No. of triangles in } \P_v(G_{\gang_v})}{\textit{No. of ``$\vee$'''s in } \P_v(G_{\gang_v})} $\\ \hline

		Group-to-group connections &  $|\{u \in \P_v(G_{\gang_v}) \textit{ s.t. } \exists (u,w) \in E$  \textit{where} $ w \notin \P_v(G_{\gang_v}) \}|$ \\\hline

		Gang-to-gang connections & $|\{u \in G_{\gang_v} \textit{ s.t. } \exists (u,w) \in E$  \textit{where} $ w \notin G_{\gang_v} \}|$\\\hline

	\end{tabular}
	
\end{table}

\begin{table}[h!]
	\caption{\textmd{Network-Based Features (Path)}}
	\label{pathFeat}
	\centering
	\renewcommand{\arraystretch}{1.5}
	
	\begin{tabular}{|p{2.5cm}|p{4.5cm}|} 
		\hline
		{\bf Description}  & {\bf Definition} \\ \hline \hline
		
		Betweenness (w.r.t. $C$)& $\sum_{u,w \in V_C}\frac{\sigma_v(u,w)}{\sigma(u,w)}$\\  \hline
		
		Closeness (w.r.t. $C$)& $(|V_C|-1)/\sum_{u\in V_C}d(u,v)$\\ \hline
		
		Shell Number (w.r.t. $C$) & $\textit{shell}_C(v)$ (see appendix for further details) \\ \hline
		
		Propagation (w.r.t. $C$) & $1$ if $v \in \Gamma_\kappa(V_\vio)$, $0$ otherwise. (see appendix for further details)\\ \hline
					
	\end{tabular}
	
\end{table}

\subsubsection{Network-Based Features}
Network-based features fall into two sub-categories that we shall describe in this section: community-based and path-based.\smallskip\\

\noindent\textbf{Network-based community features.}  We use several notions of a node's community when engineering features: the connected component to which a node belongs, the gang to which a node belongs, and what we will refer to as an individual's group.  The connected component is simply based on the overall network structure, while the gang is simply the subgraph induced by the individuals in the network who belong to the same gang (the social network of node $v$'s gang is denoted $G_{\gang_v}$.  A nodes group is defined as the partition he/she belongs to based on a partition of $G_{\gang_v}$ found using the Louvain algorithm~\cite{blondel11}.  We found in our previous work~\cite{damon13} and ensuing experience with the Chicago Police that the groups produced in this method were highly relevant operationally.  In this work, we also examined other community finding methods (i.e. \emph{Infomap}, and \emph{Spectral Clustering}) and found we obtained the best results by using the Louvain algorithm.  We provide our best performing network-based community features that we used in Table~\ref{commFeat}.  Of particular interest, we found for individual $v$ that features relating to the size of the largest connected component resulting $v'$ removal of his/her connected component was useful.  Another interesting pair of features we noted for both group and gang were the number of edges from members of that group/gang to a different group or gang.  We hypothesize that the utility of these features is a result of conflicts between groups/gangs they are connected to as well as the spread of violence amongst different groups (i.e. if two groups are closely connected, one may conduct violent activities on behalf of the other).\smallskip\\

\noindent\textbf{Network-based path features.}  We looked at several features that leveraged the paths in the network by adopting three common node metrics from the literature: betweenness, closeness~\cite{freeman77}, and shell-number~\cite{Seidman83} as well as a propagation process based on a deterministic tipping model \cite{Gran78}.  The features are listed in Table~\ref{pathFeat}.  We examined our modified definitions of closeness, betweenness, and shell number where $C$ was a single element of $\vio$, where $C=\vio$ and where $C=\emptyset$ (which provides the standard definitions of these measures).  Our intuition was that individuals nearer in the network to other violent individuals would also tend to be more violent - and we found several interesting relationships such as that for closeness (where $C=V_\vio$) discussed in section~\ref{known} when we run the classifier on each feature group. Shell number and the propagation process were used to capture the idea of the spread of violence (as shell number was previously shown to correspond with ``spreaders'' in various network epidemic models~\cite{InfluentialSpreaders_2010}).  For the propagation process, we set the threshold ($\kappa$) equal to two, three, four, five, and six.  Further details on shell number and the propogation process can be found in the appendix.\smallskip

\subsubsection{Geographic Features}
Geographic features capture the information related to the location of a crime incident. The intuition is that the individuals who commit crimes in violent districts are more likely to become violent than the others.  We found that the beat the individual has committed a crime in is an important feature for our problem. This is in accordance with previous well known literature in criminology~\cite{bb,Ros08} which studies spatio-temporal modeling of criminal behavior. The complete list is shown in Table~\ref{tab:geo_features}.
\begin{table}[!h]
	\centering
	\renewcommand{\arraystretch}{1.5}
	\caption{\textmd{Geographic Features}}
	\begin{tabular}{|p{2.5cm}|p{4.5cm}|}\hline
		\bf Name & \bf Definition \\ \hline \hline
		District Frequency & $|\{ (t, v')\ s.t.\ arr_{v'}^t = \T \land \exists t'\ s.t.\ distr_{v'}^t=distr_v^{t'} \}|$ \\ \hline
		Beat Frequency & $|\{ (t, v')\ s.t.\ arr_{v'}^t = \T \land \exists t'\ s.t.\ beat_{v'}^t=beat_v^{t'} \}|$  \\ \hline
		Beat Violence & $|\{ (t, v')\ s.t.\ arr_{v'}^t = \T \land \vio_{v'}^t \neq \emptyset \land \exists t'\ s.t.\ beat_{v'}^t=beat_v^{t'} \}|$ \\ \hline
		District Violence & $|\{ (t, v')\ s.t.\ arr_{v'}^t = \T \land \vio_{v'}^t \neq \emptyset \land \exists t'\ s.t.\ distr_{v'}^t=distr_v^{t'} \}|$  \\ \hline
	\end{tabular}
	\label{tab:geo_features}
\end{table}

\subsubsection{Temporal Features}
We considered couple of temporal features: average interval month and number of violent groups. Average interval time considers the average time duration of consecutive arrests of the offender. The other feature, which we examine, is number of violent groups appeared over time in the environment. We examined that the number of violent groups has been an important temporal aspect for identifying the violent criminals. The key intuition here is, if at least one member of the offender's groups (formed over time) is violent then we consider the offender as a part of that violent group. For an individual $v$, we define the partially ordered set $t_C^v = \{ t \st \arr_v^t = \T \land V_C^t \neq \emptyset \}$ (intuitively the set of the time points where $v$ has committed at least on of the crimes in $C$.) We also define $\Delta_i^v(C)=t_i^v-t_{i-1}^v$ for each $t_i^v \in t^v_C$.  Considering these definitions, we formally define the temporal features in Table~\ref{tab:tempo_features}.

\begin{table}[h!]
	\centering
	\renewcommand{\arraystretch}{1.5}
	\caption{\textmd{Temporal Features}}
	\begin{tabular}{|p{2.5cm}|p{4.5cm}|}\hline
		\bf Name & \bf Definition \\ \hline \hline
		
		Average interval time (w.r.t. $C$)& $ \sum\nolimits_i \Delta_i^v(C) / |t^v_C| $ \\ \hline
		
		Number of violent groups &
		\vspace{-2.8em}
		\begin{equation*}
		\begin{aligned}
		|\{ t\ s.t.\  & arr_v^t = \T \ \land \\  
		&\exists v'\ s.t.\ arr_{v'}^t=\T\ \land \\  
		& \hspace{8ex} \vio_v^t \neq \emptyset\ \land \\
		& \hspace{8ex} v' \in N_{v}^t \}|
		\end{aligned}
		\end{equation*} \\ \hline
		
	\end{tabular}
	\label{tab:tempo_features}
\end{table}
\section{Experimental Results}
\label{sec:exp_res}
In this section, we review the results of our experiments.  We looked at two types: experiments where the entire co-offender network is known before-hand (Section~\ref{known}) and experiments where the network is discovered over time (Section~\ref{unknown}).  The intuition behind the experiments where the co-offender network is known is that the police often have additional information to augment co-arrestee data.  This information can include informant reporting, observed individuals interacting by patrolmen, intelligence reporting, and information discovered on social media and the Internet.  In our second type of experiment we discover the network over time in an effort to mimic real-world operations - however, we also show that this makes the problem more difficult as it reduces the power of neighborhood-based and network-based features.  Based on our discussions with the Chicago Police, we believe that real-world results will most likely fall somewhere between these two experiments.  Operationally, we will not have full arrest data, but the aforementioned augmenting data sources are available (even though we did not have access to them for our experiments).

\subsection{Known Co-Offender Network}
\label{known}
In this experiment we assume that the entire offender network is known. In other words, to compute the features for each vertex $v$, we assume that the set $\vio_v$ is unknown while the rest of the network is observable.  In here we compared our approach with \emph{THH} but not with the \emph{PVA} as we do not utilize time.  In each of the experiments described in this section, we conduct $10$-fold cross validation.  We consider the result of each approach as a set of nodes that the approach considers to be a set of potentially violent individuals.  Our primary metrics are precision (fraction of reported violent individual who were actually violent in the dataset), recall (fraction of violent individuals in the dataset reported by the approach), F1 (the harmonic mean of precision and recall) and area under the curve.  We conduct two types of experiments: first, we study classification performance using only features within a given category (neighborhood, network, temporal, and geographic), then we study the classification performance when the entire feature set is used but with various different classification algorithms and compare the result to \emph{THH}.\smallskip\\

\begin{figure}[h!]
	\centering
	\includegraphics[width=7cm]{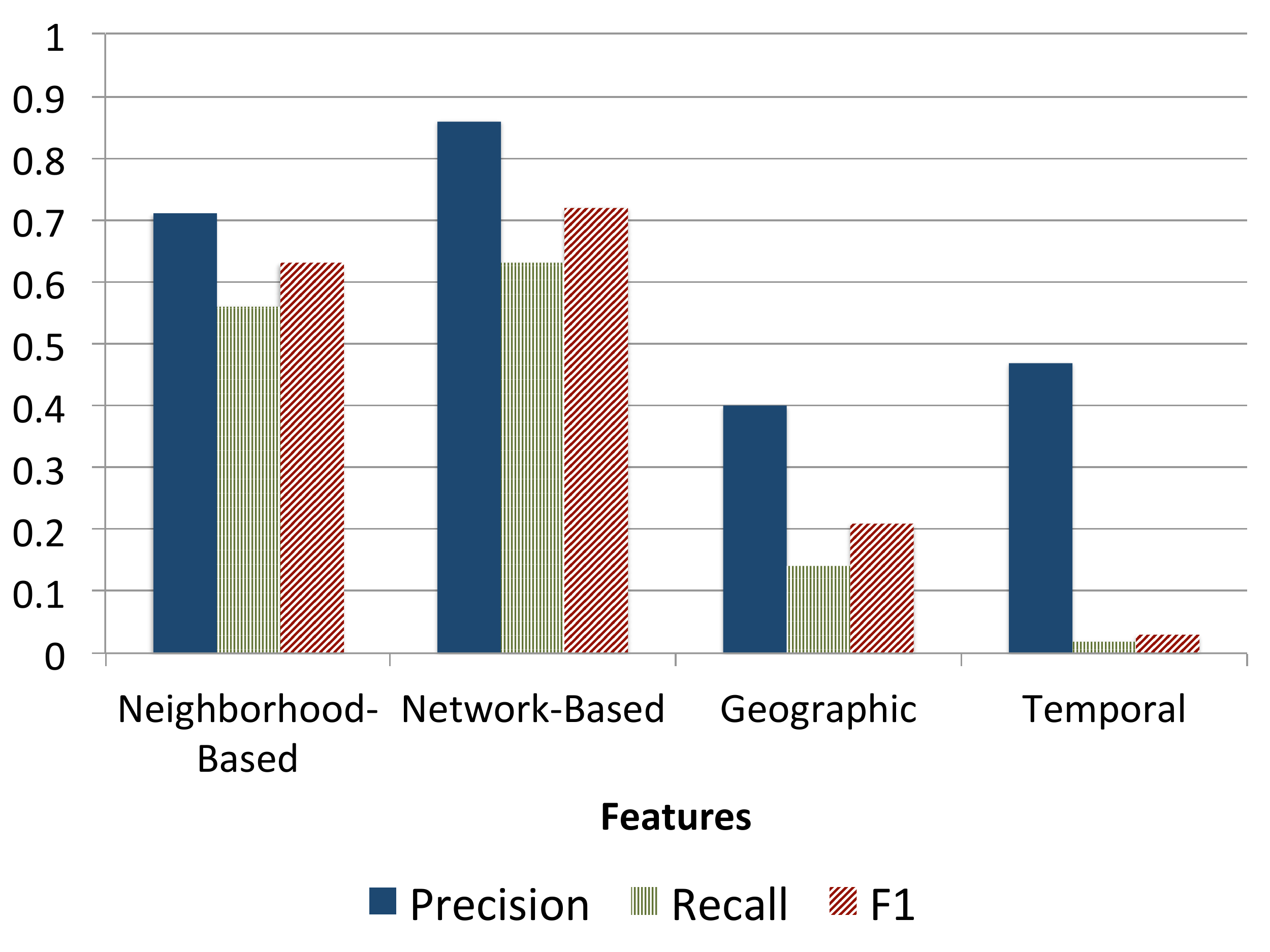}
	\caption{\textmd{Precision, recall, and F1 comparison between each group of features.}}
	\label{fig:feat_cpmp}
\end{figure}

\begin{figure}[h!]
	\centering
	\subfigure[] {
		\epsfig{figure=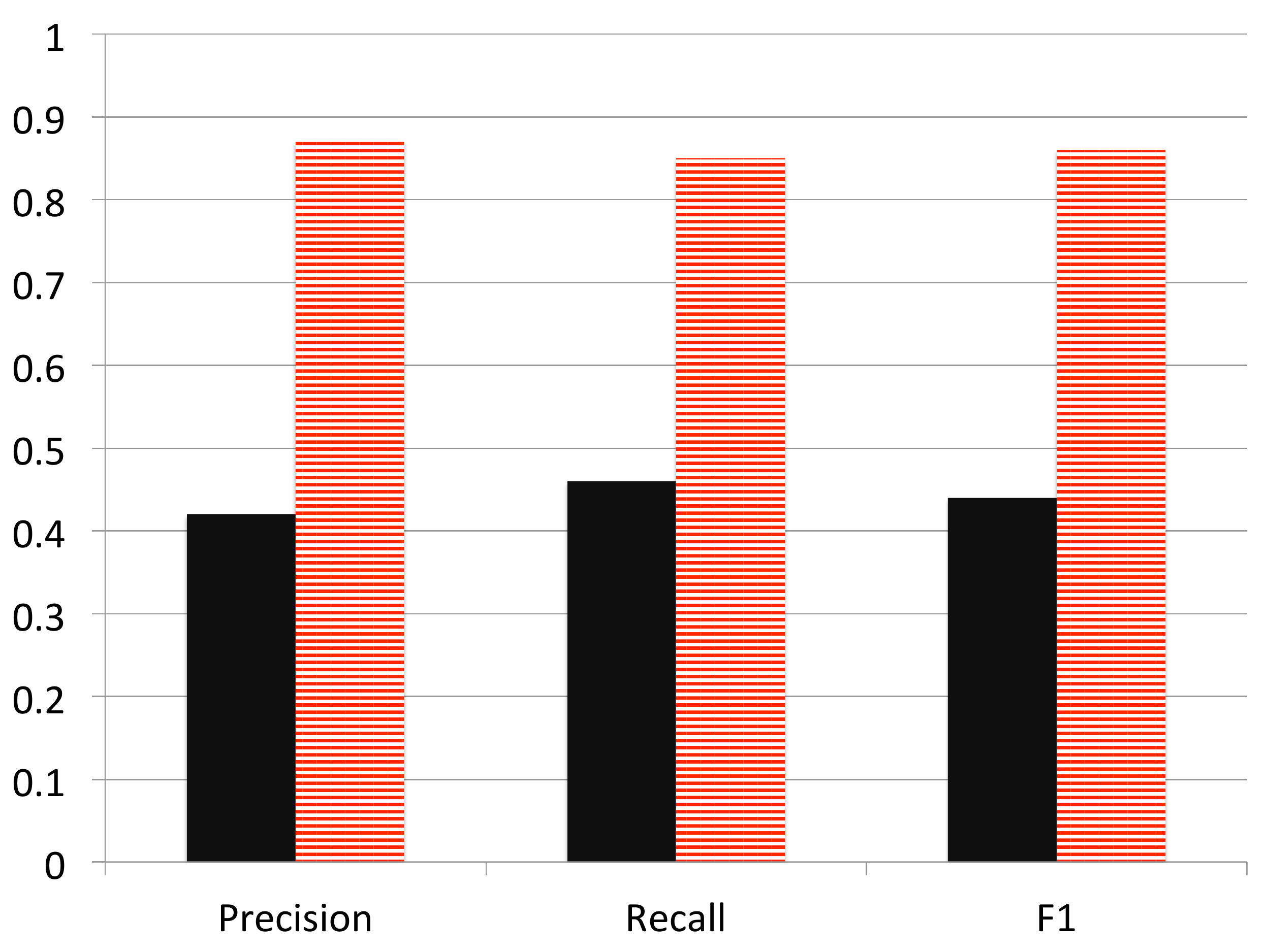,width=3.7cm}
	}
	\quad
	\subfigure[] {
		\epsfig{figure=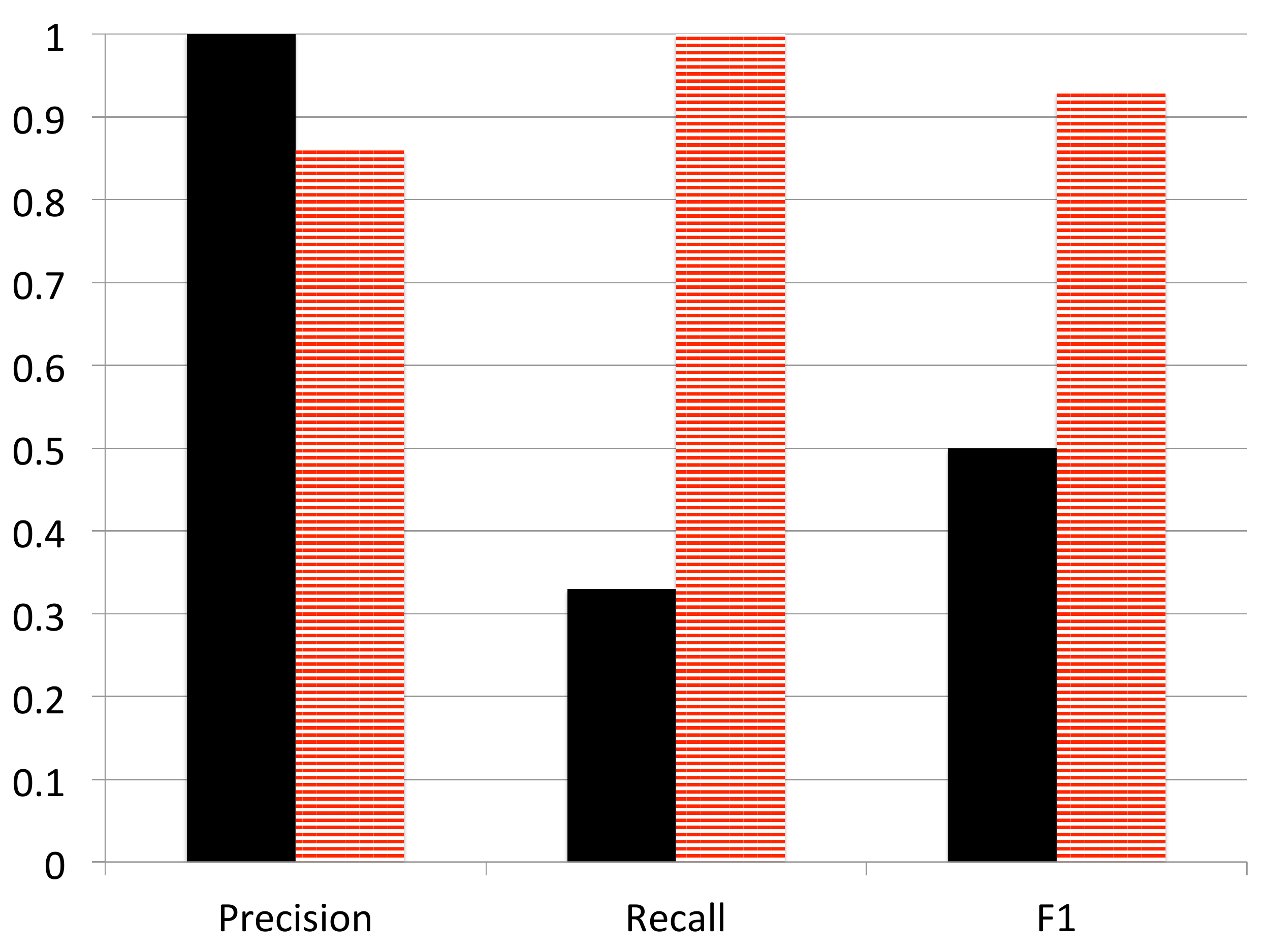,width=3.7cm}
	}
	\quad
	\subfigure[] {
		\epsfig{figure=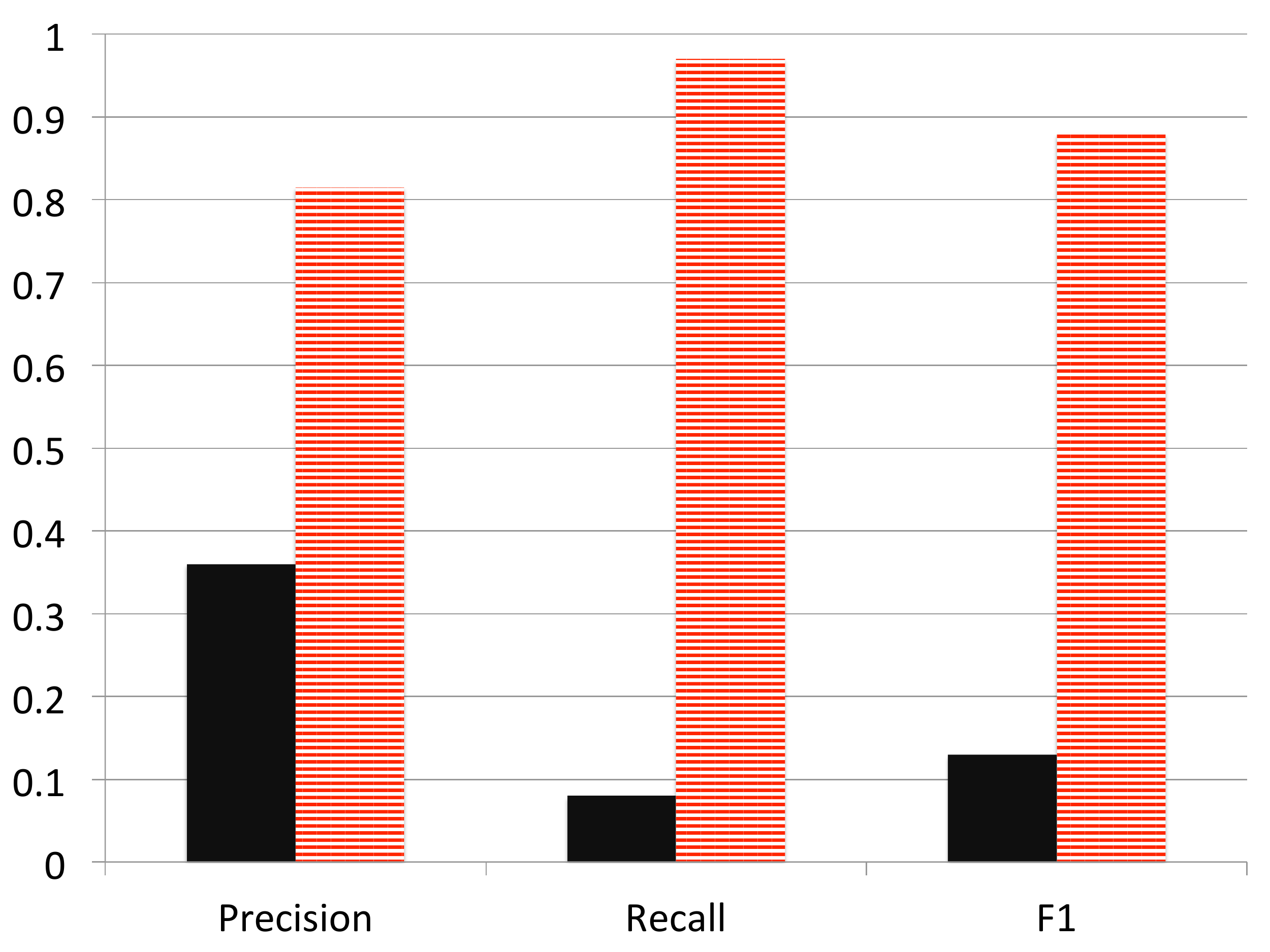,width=3.7cm}
	}
	\quad
	\subfigure[] {
		\epsfig{figure=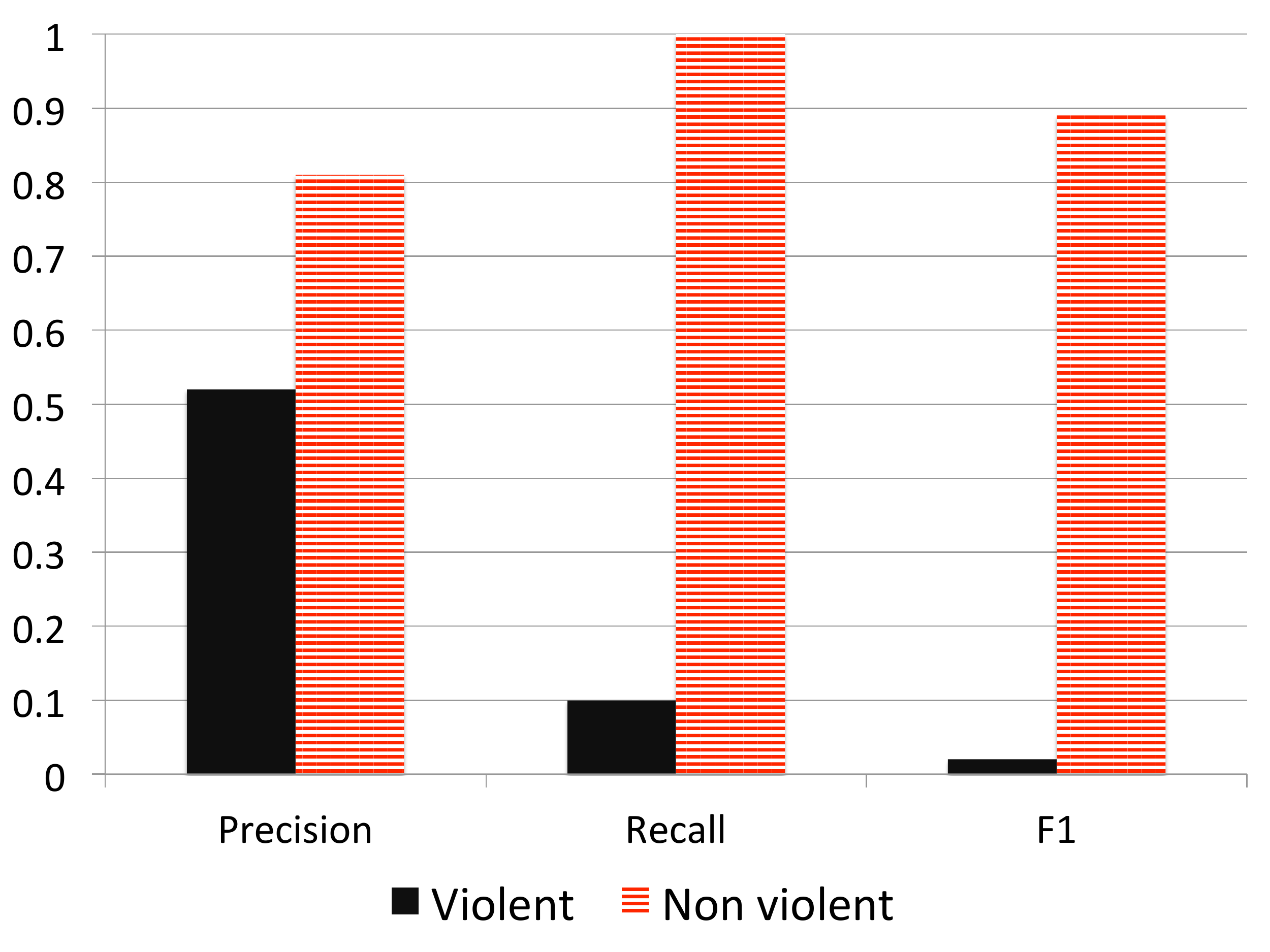,width=3.7cm}
	}
	\caption{Example features from each category. (a) Neighborhood-based: Minority of 1-hop and majority of 2-hop neighbors committing a crime in $C$. (b) Network-based: Closeness (w.r.t. $\vio$). (c) Geographic: Beat violence. (d) Temporal: Average interval months.}
	\label{fig:inv_feat}
\end{figure}

\noindent\textbf{Classification using single feature categories.}  Here we describe classification results using single feature categories.  In this set of experiments, we use a random forest classifier (which we will later show provides the best performance of the classifiers that we examined). Figure~\ref{fig:feat_cpmp} shows the performance of RF for  the described categories. The network-based features are highly-correlated to violent behavior with average F1 value of 0.72 compared to 0.63 for neighborhood, 0.21 for geographic, and 0.03 for temporal features. In Figure~\ref{fig:inv_feat}, we show the performance of a feature from each category to classify violent vs. non violent crimes; the performance of each example is a good indicator of the performance of its category.\smallskip\\

\begin{figure}[h!]
	\centering
	\includegraphics[width=7cm]{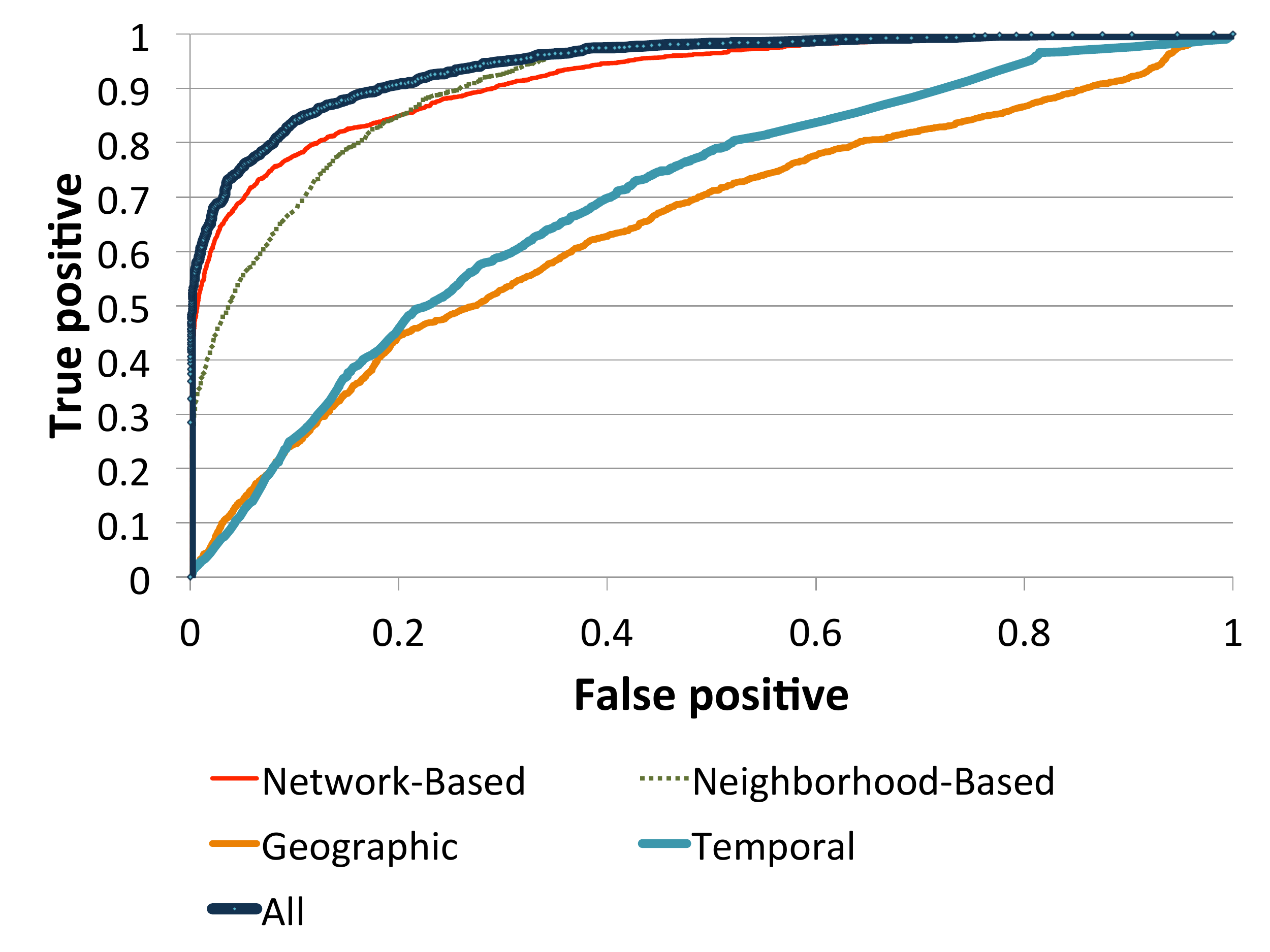}
	\caption{ROC curve for each feature set.}
	\label{fig:roc}
\end{figure}

\begin{figure}[h!]
	\centering
	\includegraphics[width=7cm]{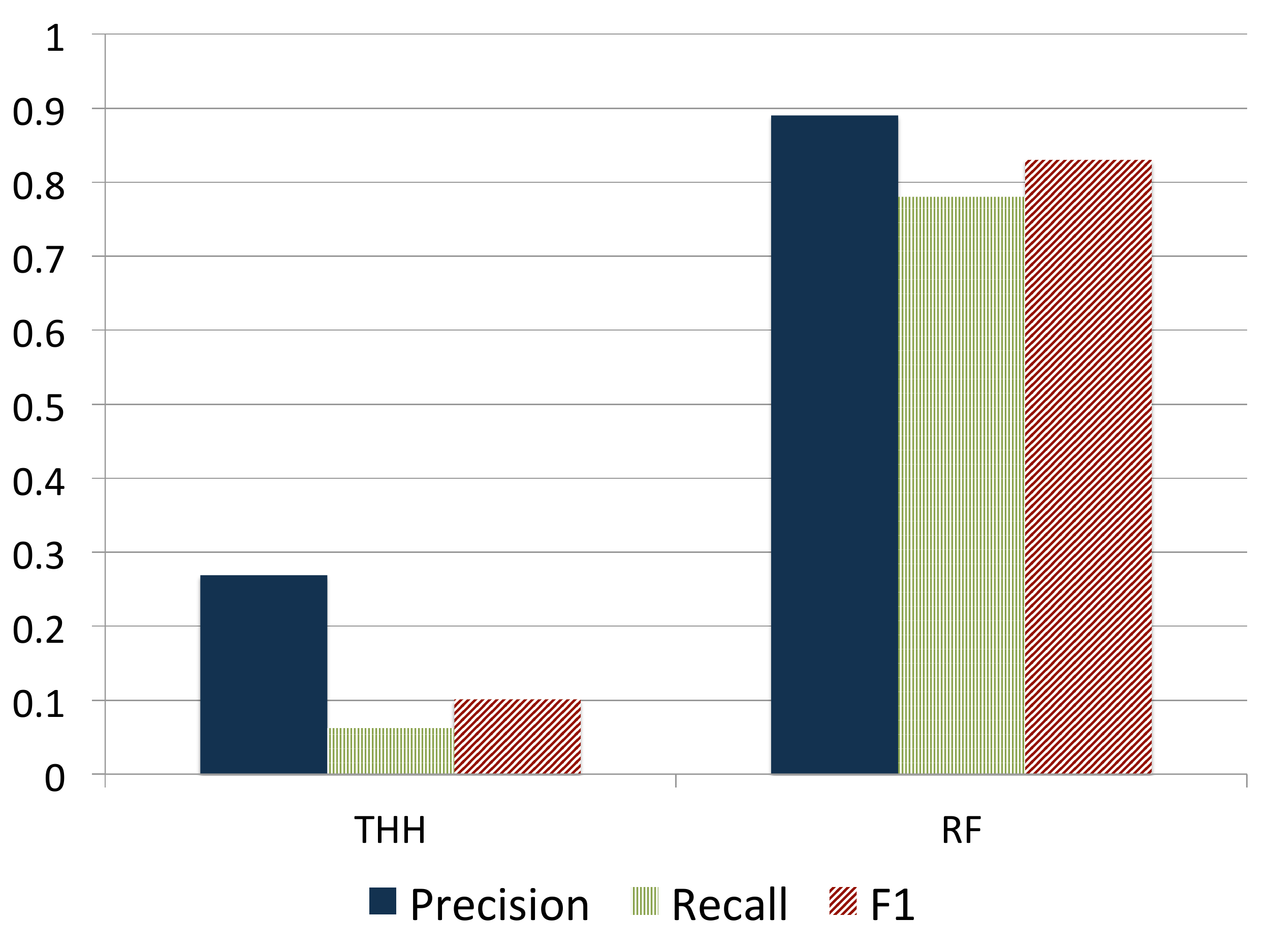}
	\caption{\textmd{Performance comparison between \emph{THH} and \emph{RF} in K-fold cross validation.}}
	\label{fig:compKf}
\end{figure}

\noindent\textbf{Classification comparison.}  Table~\ref{tab:kfold_all} shows the performance of different classification algorithms. According to Table~\ref{tab:kfold_all}, RF provides the best performance (F1=0.83); we also note that using SMOTE for RF, did not improve this result. Figure~\ref{fig:compKf} shows that our algorithm outperforms \emph{THH}. The performance of our features are also illustrated in Figure~\ref{fig:roc}. The area under the curve (AUC) of applying all features is 0.98 -- a higher overall accuracy. The AUC for network-based, neighborhood-based, geographic, and temporal categories are 0.92, 0.91, 0.65, and 0.7 respectively.  This indicates the importance of network features for this classification task.\\

\begin{table}
	\centering
	\caption{\textmd{K-fold cross validation.}}
	\begin{tabular}{| l | l l l |}
		\hline
		Method &  Precision &  Recall & F1 \\ \hline
		\hline
		RF & 0.89 & 0.78 & 0.83\\ \hline
		RF w. SMOTE & 0.86 & 0.78 & 0.82 \\ \hline
		NB & 0.45 & 0.49 & 0.47 \\ \hline
		LR & 0.68 & 0.49 & 0.57 \\ \hline
		DT & 0.71 & 0.66 & 0.68 \\ \hline
		NN & 0.64 & 0.57 & 0.6 \\ \hline
		SVM & 0.73 & 0.2 & 0.31 \\ \hline
	\end{tabular}
	\label{tab:kfold_all}
\end{table}

\subsection{Co-Offender Network Emerges Over Time}
\label{unknown}
\begin{figure}[h!]
	\centering
	\epsfig{figure=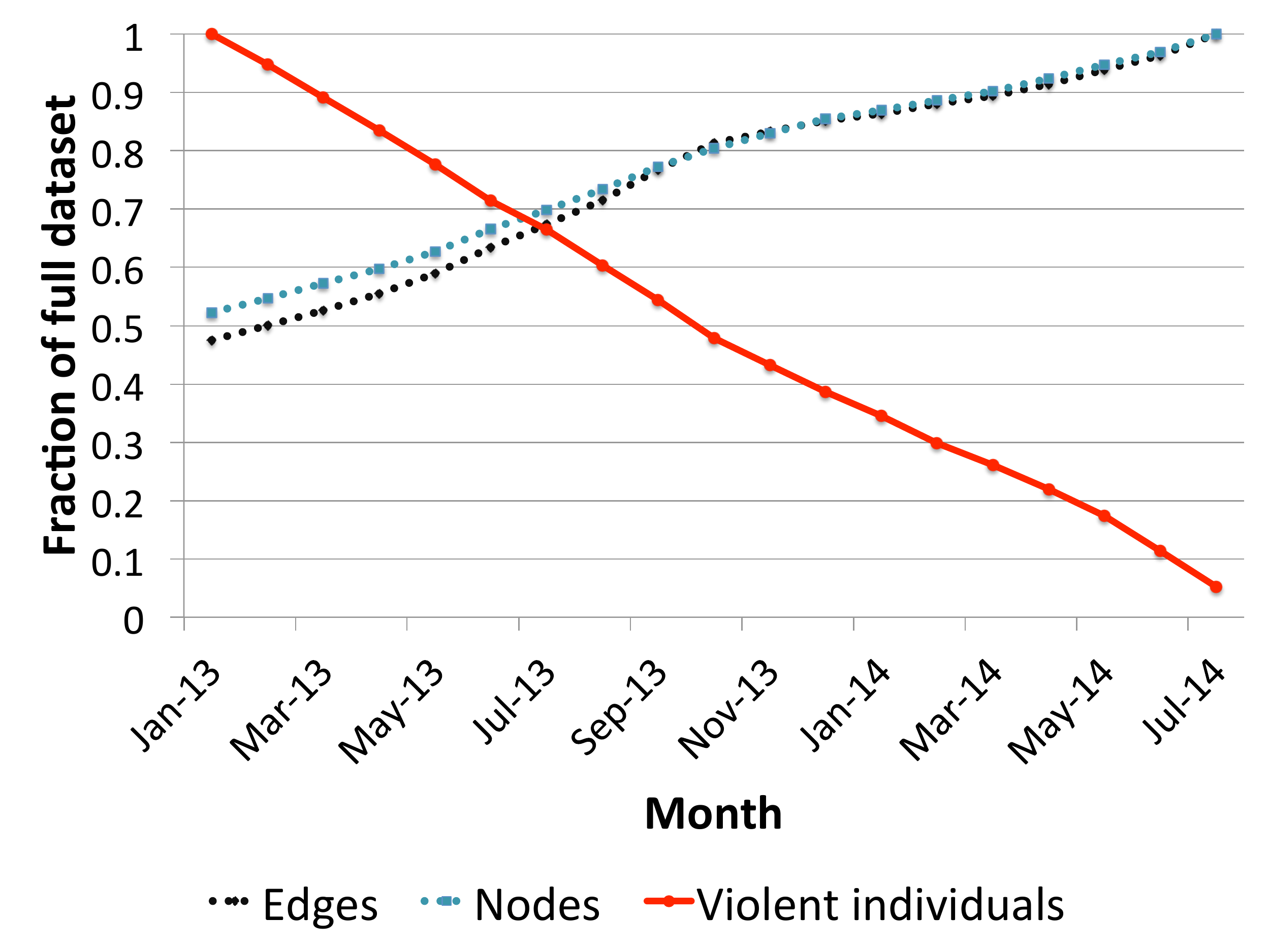,width=7cm}
	\caption{\textmd{Number of nodes, edges, and violent individuals over time. More training data, less offenders to identify.}}
	\label{graphEvol}
\end{figure}

In this section, we present a more difficult experiment - where the co-arrestee network is discovered over time (by virtue of arrests).  To simulate this phenomenon, we split our data into two disjoint sets: the first set for learning and identification, and the second one for measuring the performance. We do monthly split and start from February 2013.  To illustrate the difficulty of this test, we show the number of nodes, edges, and violent individuals per month in Figure~\ref{graphEvol}.  We note that in the early months, we are missing much of the graphical data (over $40\%$ of nodes and edges in the first two months) - hence making many of our features less effective.  However, as the months progress, there are less violent individuals to identify (due to the temporal nature of the dataset) - hence amplifying the data imbalance as time progresses.\smallskip

\begin{figure}[h!]
	\centering
		\subfigure {
			\epsfig{figure=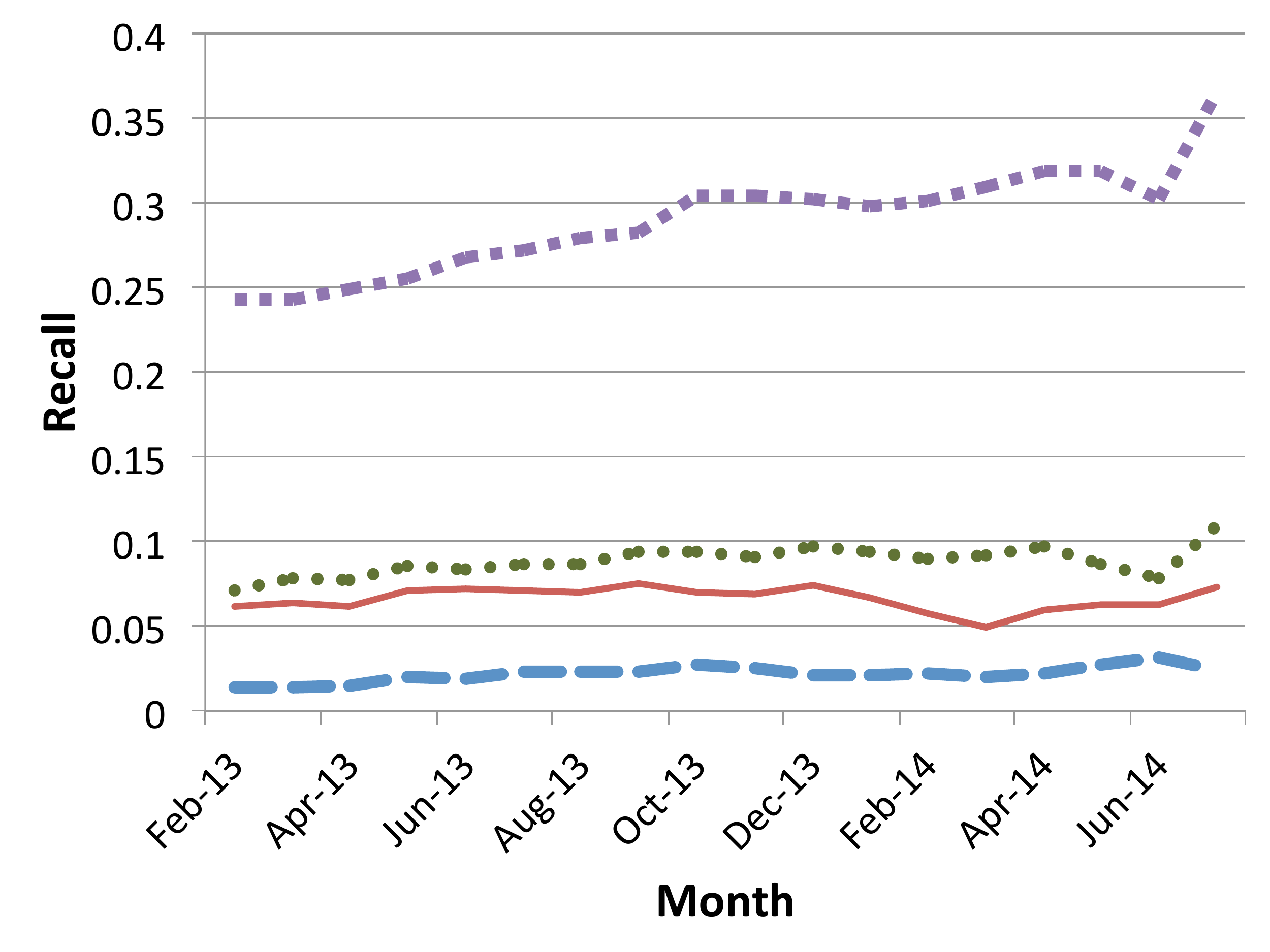,width=7cm}
		}
		\quad
		\subfigure {
			\epsfig{figure=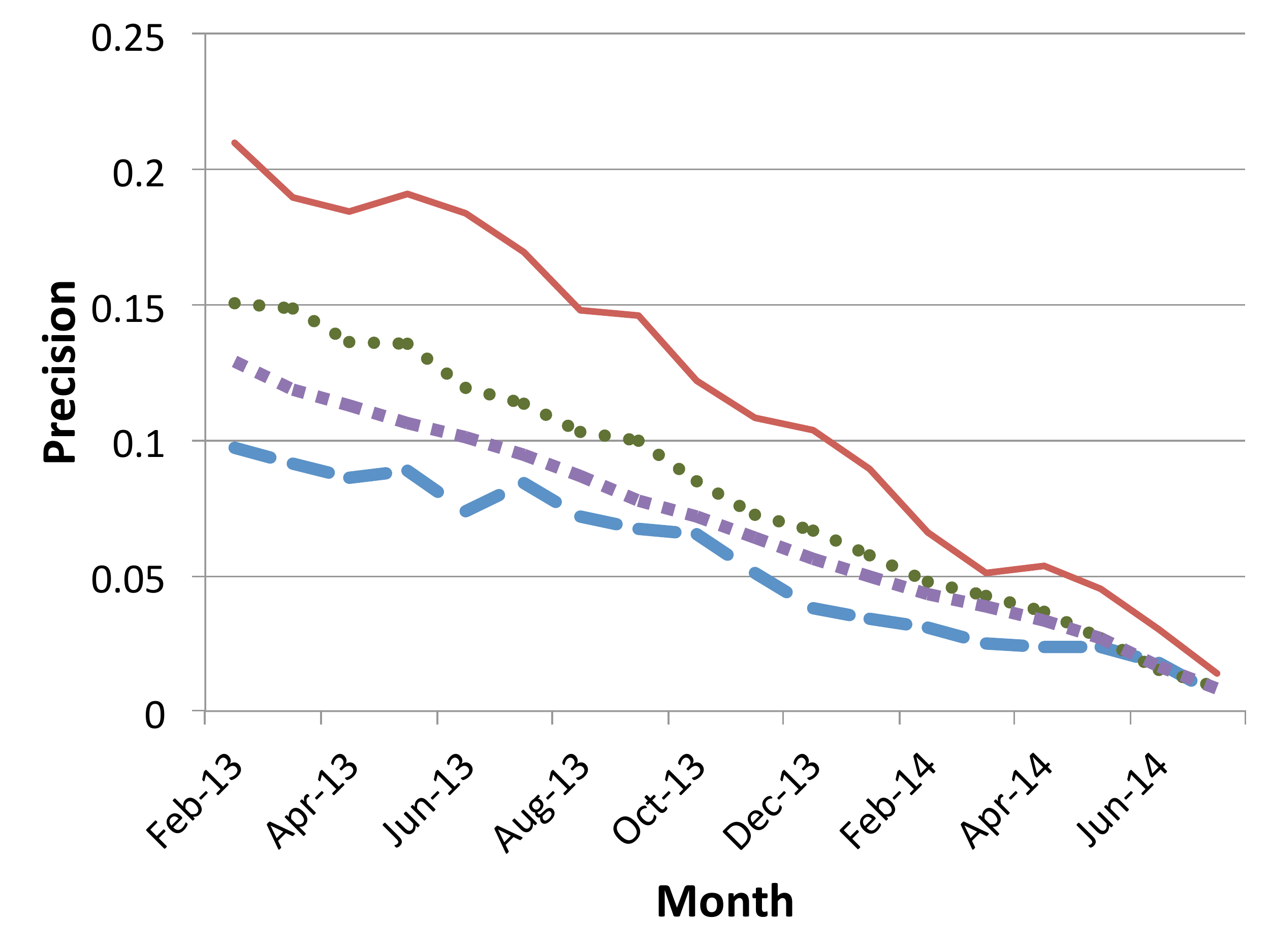,width=7cm}
		}
		\quad
		\subfigure {
			\epsfig{figure=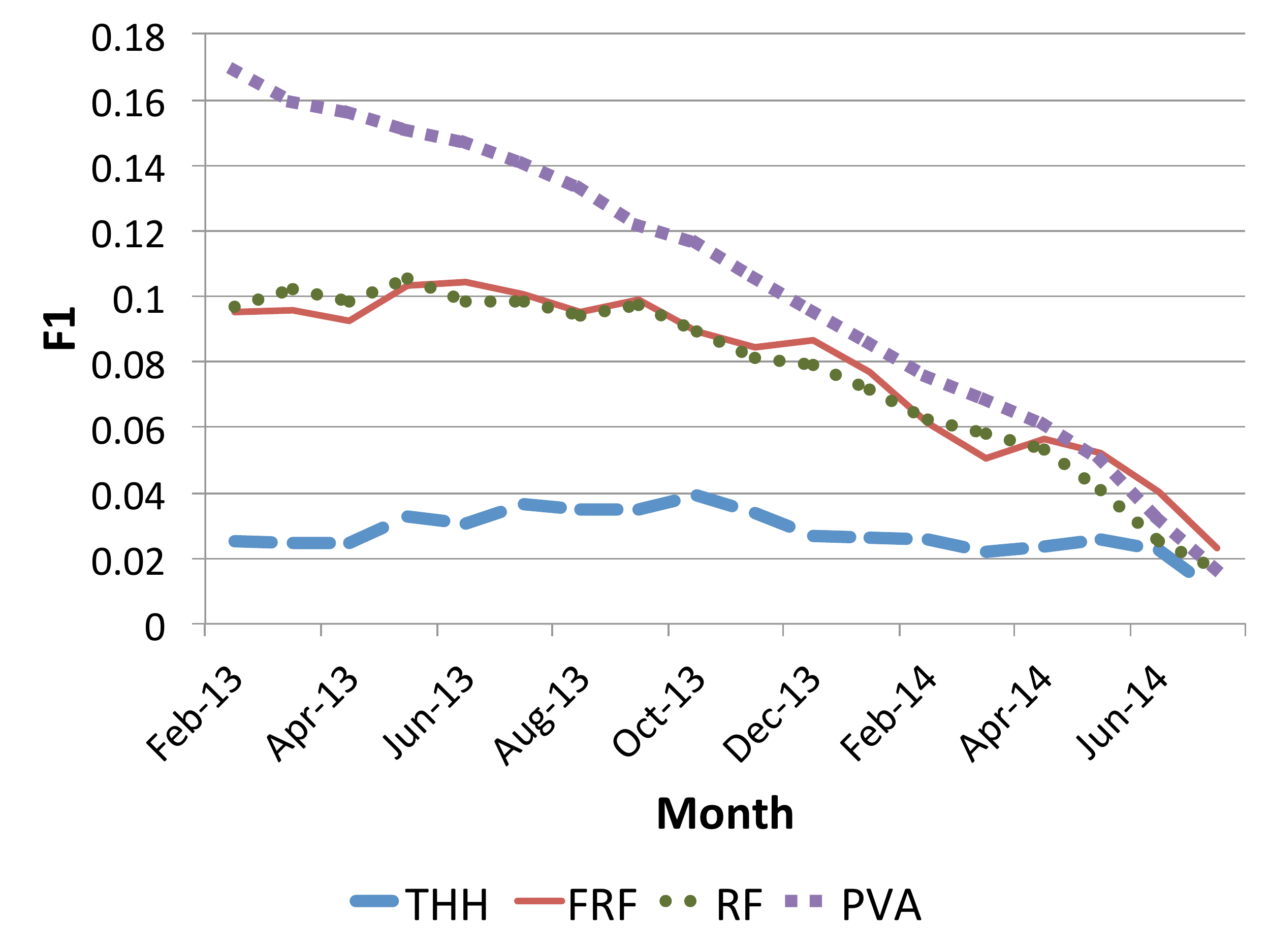,width=7cm}
		}
	\caption{\textmd{Precision, recall, and F1 over time.}}
	\label{fig:time_test_prf}
\end{figure}

\begin{figure}[h!]
	\centering
		\subfigure {
			\epsfig{figure=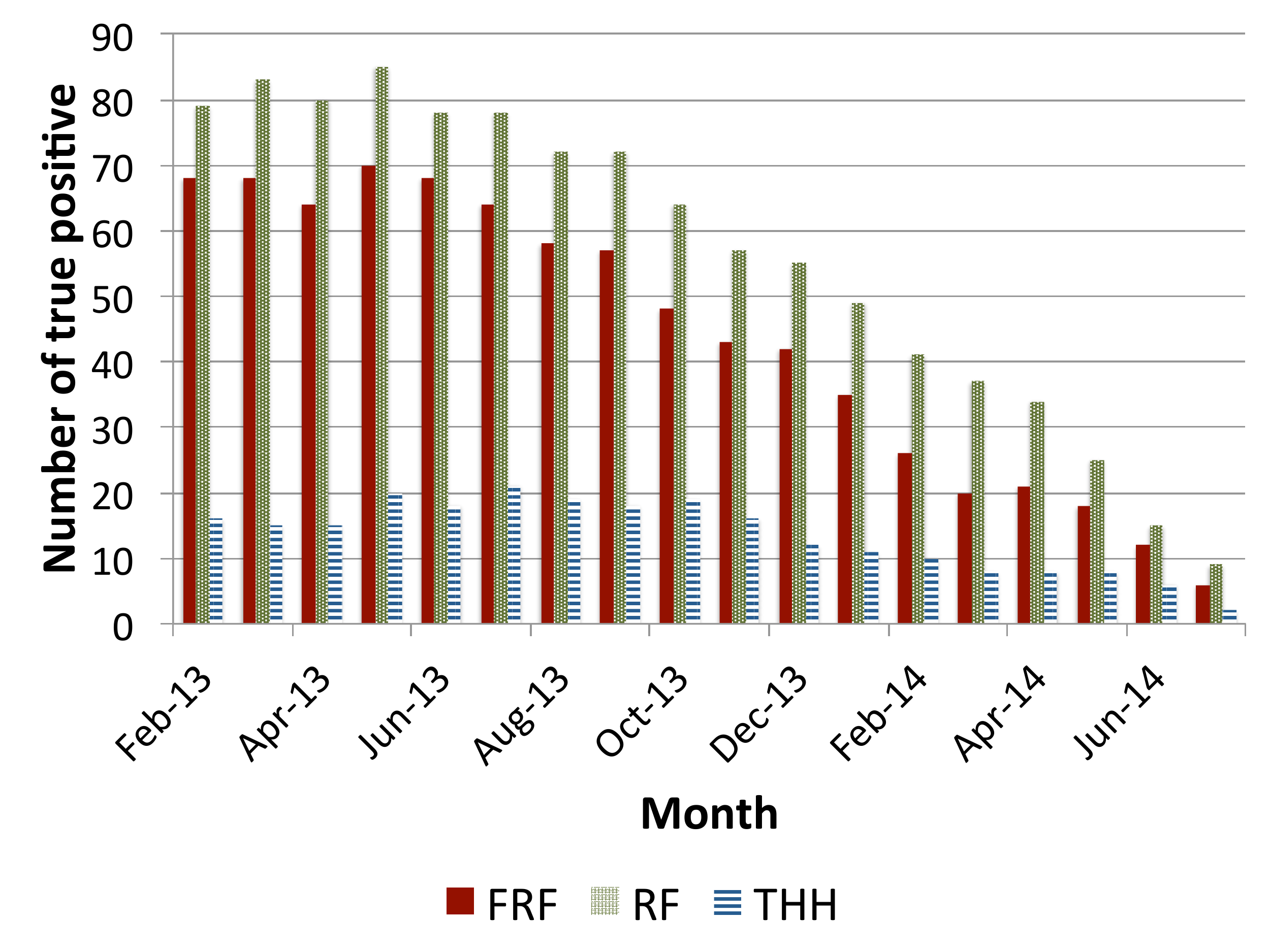,width=7cm}
		}
		\quad
		\subfigure {
			\epsfig{figure=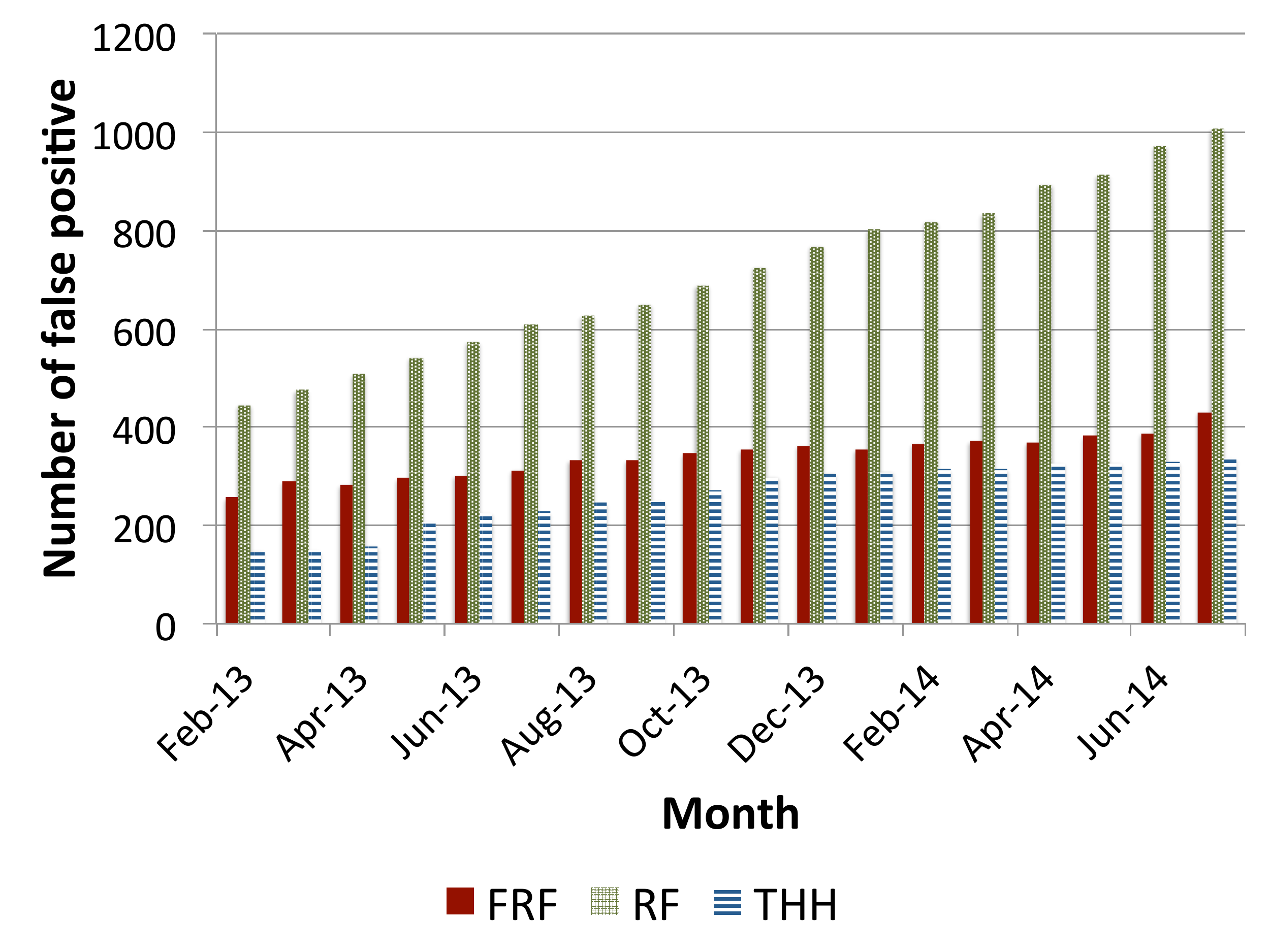,width=7cm}
		}
	\caption{\textmd{Number of true and false positive instances.}}
	\label{fig:comp}
\end{figure}

In these experiments, we compared our approach using random forests with the full feature set to \emph{THH} and \emph{PVA}.  We measure precision, recall, F1, number of true positives, and number of false positives and display the results in Figures~\ref{fig:time_test_prf} and \ref{fig:comp}. In \emph{FRF} (Filtered Random Forest) we filter the offenders who have not committed any crime in the last 200 days. This simple heuristic increase the precision drastically while preserving the recall. The main advantage of our method, besides the high precision, is its ability to significantly reduce the population of potentially violent offenders when compared to PVA - which for each month had between $1813$ and $3571$ false positives.  Figure~\ref{fig:comp} compares the number of true and false positives instances for all the approaches for each month except PVA (PVA was omitted due to readability because of the large amount of false positives). While the F1 measure for \emph{PVA} is higher than that of the others, the large number of false positives prevents the law enforcement from using it effectively in practice.  Furthermore, as time progresses, PVA likely rises in recall due to the drop in the number of violent criminals to predict.

\section{Related Work}
\label{rwSec}

Though we believe that the prediction of violent offenders using co-offender social networks is new, there has previously been work on both co-offender networks in general as well as crime forecasting.  In this section, we briefly review some of the relevant contributions in both of these areas.\smallskip

There has been much previous work on co-offender networks.  The earlier work that studied these special social networks primarily came from the criminology literature.  For instance, \cite{mors09} utilizes social network analysis techniques to study several case studies where the social network of the criminal organization was known.  In \cite{mcg08}, the authors study the stability of these networks change over time.  More recently, graphical features derived from networks comprised of both offenders and victims has been shown to be related to the the probability of an individual becoming a victim of a violent crime~\cite{pap12,pap15}.  Previous work has also looked at the relationship between network structure and geography~\cite{pap13} and has leveraged both network and geographic features to predict criminal relationships~\cite{Tayebi14} as well as influence gang members to dis-enroll~\cite{shak14}.  There have also been several software tools developed for conducting a wide-range of analysis on co-offender networks including  CrimeFighter~\cite{conf/eisic/PetersenW11}, CrimeLink~\cite{sch11}, and ORCA~\cite{damon13}.  However, our work departs from this is that we are looking to leverage the network topology and other features to identify violent offenders - which was not studied in any of the previous work.\smallskip

There has also been a large amount of work on crime forecasting (i.e. \cite{gorr,liu03}) though historically, this work has relied on spatio-temporal modeling of criminal behavior~\cite{bb,Ros08} or was designed to identify suspects for specific crimes~\cite{tay11,Hammer}.  None of this previous work was designed to identify future violent offenders nor did it leverage social network structure. \\ 
\section{Conclusion}
In this paper we explored the problem of identifying repeat offenders who will commit violent crime.  We showed a strong relationship between network-based features and whether a criminal will commit a violent offense providing an unbiased F1 score of $0.83$ in our cross-validation experiment where we assumed that the underlying network was known.  When we moved to the case where the network was discovered over time, our method significantly outperformed baseline approaches significantly increasing precision and recall.  We are currently discussing ways to operationalize this technology with the Chicago Police as well as design strategies to best deploy police assets to areas with higher concentrations of potentially violent offenders.  We are also working with the police to identify other sources of data to build a more complete social network of the offenders.

\section*{Appendix}
\noindent\textbf{Shell Number.}  For a given graph, the $k$-core is the largest subgraph where each node has at least degree $k$.  The $k$-shell is the set of nodes in core $k$ but not in any higher core.  A node's shell number is $k$ value of the shell to which that node belongs.  For a given node $v$ and $C \subseteq \vio$, we define $\textit{shell}_C(v)$ as the shell number of node $v$ on the subgraph consisting of $v$ and all nodes $v'$ where $C \cap \vio_v \neq \emptyset$.  We slightly abuse notation and define $\textit{shell}_\emptyset(v)$ as the shell number of $v$ on the full network.\smallskip\\

\noindent {\textbf{Propogation Process.}}  For a given node $v$ and the set of activated nodes $V'$, we define $v$'s active neighbors as follows:
\begin{equation*}
act_v(V') = \{u | u \in N_v^1  \cap V'\}
\end{equation*}
We now define an activation function $A$ that, given an initial set of active nodes, returns a set of active nodes after one time step.
\begin{equation*}
A_{\kappa}(V')=V' \cup \{ v \in V \textit{ s.t. } |act_v(V')| \geq \kappa \}
\end{equation*}

We also note that the activation function can be applied iteratively, to model a diffusion process. Hence, we shall use the following notation to signify multiple applications of A (for natural numbers $t > 1$).
\begin{equation*}
A_{\kappa}^t(V') =  \left \{
\begin{array}{l l}
A_{\kappa}(V')& \text{if} \ t=1 \\
A_{\kappa}(A_{\kappa}^{t-1}(V') )& \text{otherwise}
\end{array}
\right.\ 
\end{equation*}

Clearly, when $A_{G,\kappa}^t(V') =  A_{G,\kappa}^{t-1}(V')$ the process has converged. Further, this always converges in no more than $|V|$ steps, since the process must activate at least one new node in each step prior to converging. Based on this idea, we define the function $\Gamma$ which returns the set of all nodes activated upon the convergence of the activation function.  We define $\Gamma_\kappa(V')=A_{\kappa}(V')$ where $t$ is the least value such that $A_{\kappa}^t(V') =  A_{\kappa}^{t-1}(V')$.


\end{document}